\documentclass[twocolumn, floatfix]{aastex631}
\usepackage{ulem}
\usepackage{float}
\usepackage{gensymb}

\usepackage{breqn}
\usepackage[flushmargin]{footmisc}

\usepackage{amsmath}
\newcommand{\angstrom}{\textup{\AA}}

\shorttitle{Phase-space Properties and Chemistry of Sagittarius Stream}
\shortauthors{Limberg et al.}

\graphicspath{{./}{figures/}}

\begin{document}

\title{%The SEGUE/\texttt{StarHorse} View of the Ancient Milky Way. I. \\%Correlations Between 
Phase-space Properties and Chemistry of the Sagittarius Stellar Stream Down to \\ the Extremely Metal-poor ($\rm[Fe/H] \lesssim -3$) Regime
}

\correspondingauthor{Guilherme Limberg}
\email{guilherme.limberg@usp.br}

\author[0000-0002-9269-8287]{Guilherme Limberg}
\affil{Universidade de S\~ao Paulo, Instituto de Astronomia, Geof\'isica e Ci\^encias Atmosf\'ericas, Departamento de Astronomia, \\ SP 05508-090, S\~ao Paulo, Brasil}
\affil{Department of Astronomy \& Astrophysics, University of Chicago, 5640 S. Ellis Avenue, Chicago, IL 60637, USA}
\affil{Kavli Institute for Cosmological Physics, University of Chicago, %5640 S. Ellis Avenue, 
Chicago, IL 60637, USA}

\author[0000-0001-9209-7599]{Anna B. A. Queiroz}
\affil{Leibniz-Institut f$\ddot{u}$r Astrophysik Potsdam (AIP), An der Sternwarte 16, 14482 Potsdam, Germany}
\affil{Institut f\"{u}r Physik und Astronomie, Universit\"{a}t Potsdam, Haus 28 Karl-Liebknecht-Str. 24/25, 14476 Golm, Germany}

\author[0000-0002-0537-4146]{H\'elio D. Perottoni}
\affil{Universidade de S\~ao Paulo, Instituto de Astronomia, Geof\'isica e Ci\^encias Atmosf\'ericas, Departamento de Astronomia, \\ SP 05508-090, S\~ao Paulo, Brasil}
\affil{Institut de Ciències del Cosmos (ICCUB), Universitat de Barcelona (IEEC-UB), Martí i Franquès 1, E08028 Barcelona, Spain}

\author[0000-0001-7479-5756]{Silvia Rossi}
\affil{Universidade de S\~ao Paulo, Instituto de Astronomia, Geof\'isica e Ci\^encias Atmosf\'ericas, Departamento de Astronomia, \\ SP 05508-090, S\~ao Paulo, Brasil}

\author[0000-0002-7662-5475]{Jo\~ao A. S. Amarante}\altaffiliation{Visiting Fellow at UCLan}
\affiliation{Institut de Ciències del Cosmos (ICCUB), Universitat de Barcelona (IEEC-UB), Martí i Franquès 1, E08028 Barcelona, Spain}
\affil{Jeremiah Horrocks Institute, University of Central Lancashire, Preston, PR1 2HE, UK}

\author[0000-0002-7529-1442]{Rafael M. Santucci}
\affiliation{Universidade Federal de Goi\'as, Instituto de Estudos Socioambientais, Planet\'ario, Goi\^ania, GO 74055-140, Brasil}
\affiliation{Universidade Federal de Goi\'as, Campus Samambaia, Instituto de F\'isica, Goi\^ania, GO 74001-970, Brasil}

\author[0000-0003-1269-7282]{Cristina Chiappini}
\affil{Leibniz-Institut f$\ddot{u}$r Astrophysik Potsdam (AIP), An der Sternwarte 16, 14482 Potsdam, Germany}
\affil{Laborat\'orio Interinstitucional de e-Astronomia - LIneA, RJ 20921-400, Rio de Janeiro, Brasil}

\author[0000-0002-5974-3998]{Angeles P\'erez-Villegas}
\affil{Instituto de Astronom\'ia, Universidad Nacional Aut\'onoma de M\'exico, Apartado Postal 106, C. P. 22800, Ensenada, B. C., M\'exico}

\author[0000-0001-5297-4518]{Young Sun Lee}
\affiliation{Department of Astronomy and Space Science, Chungnam National University, Daejeon 34134, Republic of Korea}
\affiliation{Department of Physics and Astronomy and JINA Center for the Evolution of the Elements, University of Notre Dame, \\ Notre Dame, IN 46556, USA}

\begin{abstract}
%Sagittarius (Sgr) dwarf spheroidal is the textbook example of a satellite galaxy undergoing tidal stripping and its associated debris %cover a full circle across the sky and 
%are referred to as ``Sgr stream''. 
In this work, we study the phase-space and chemical properties of Sagittarius (Sgr) stream, the tidal tails produced by the ongoing destruction of Sgr dwarf spheroidal (dSph) galaxy, %but 
focusing on its very metal-poor (VMP; $\rm[Fe/H] < -2$) content. %For the task, 
We combine spectroscopic and astrometric information from SEGUE and \textit{Gaia} EDR3, respectively, with data products from a new large-scale run of \texttt{StarHorse} spectro-photometric code. Our selection criteria yields ${\sim}1600$ stream members, including ${>}200$ VMP stars. We find the leading arm ($b>0^{\circ}$) of Sgr stream to be more metal-poor, by ${\sim}0.2$\,dex, than the trailing one ($b<0^{\circ}$). With a subsample of turnoff and subgiant stars, we estimate this substructure's stellar population to be ${\sim}1$\,Gyr older than the thick disk's. With the aid of an $N$-body model of the Sgr system, we verify that simulated particles stripped earlier (${>}2$\,Gyr ago) have present-day phase-space properties similar to lower-metallicity stream stars. Conversely, those stripped more recently (${<}2$\,Gyr) are preferentially more akin to metal-rich ($\rm[Fe/H] > -1$) members of the stream. Such correlation between kinematics and chemistry can be explained by the existence of a dynamically hotter, less centrally-concentrated, and more metal-poor population in Sgr dSph prior to its disruption, implying that this galaxy was able to develop a metallicity gradient before its accretion. Finally, we identified several carbon-enhanced metal-poor ($\rm[C/Fe] > +0.7$ and $\rm[Fe/H] \leq -1.5$) stars in Sgr stream, which might be in tension with current observations of its remaining core where such objects are not found.
\end{abstract}

\section{Introduction} \label{sec:intro}
\setcounter{footnote}{13}

%The halos of massive galaxies are assembled through a succession of merging events
The Galactic stellar halo is expected to be assembled through a succession of merging events between the Milky Way and dwarf galaxies of various masses in the context of the hierarchical formation paradigm \citep{sz1978, White1991cdm, Kauffmann1993, Springel2006}. Upon interacting with the Galactic gravitational potential well, the constituent stars of these satellites become tidally unbound and, over time, phase-mixed into a smooth halo \citep[e.g.,][]{HelmiWhite1999}. The intermediate stage of this process is characterized by the appearance of stellar streams, spatially elongated structures produced by accreted debris that remains kinematically cohesive \citep{Johnston1998, BullockJohnston2005, Cooper2010, Cooper2013, Pillepich2015halos, Morinaga2019_MWhalos}. 

The magnificence of immense stellar streams can be appreciated both in external massive galaxies (e.g., M31/Andromeda, NGC~5128/Centaurus A, and M104/Sombrero; \citealt{Ibata2001streamM31}, \citealt{Crnojevic2016_streamsCenA}, \citealt{MartinezDelgado2021sombrero}) as well as in the Milky Way itself (e.g., \citealt{Belokurov2006Streams}). The archetype of the above-described process is the Sagittarius (Sgr) stream \citep[e.g.,][]{Mateo1998}, the tidal tails produced by the destruction of Sgr dwarf spheroidal (dSph) galaxy \citep{Ibata1994, Ibata1995}.  

Over the past couple of decades, the Sgr stream has been mapped across ever increasing areas of the sky \citep{Mateo1996, Mateo1998, Alard1996sgr, Ibata2001sgrStream, Newberg2003SgrStream, MartinezDelgado2004sgr}. Eventually, wide-area photometric data %from the Two Micron All Sky Survey (2MASS; \citealt{2MASS} and references therein) 
allowed us to contemplate the grandiosity of Sgr stream throughout both hemispheres \citep{Majewski2003, Yanny2009sgr}. Furthermore, observations of distant halo tracers \citep[e.g., RR Lyrae stars;][]{Vivas2005sgrRRL} and line-of-sight velocity ($v_{\rm los}$) measurements \citep[][]{Majewski2004sgr} served as constraints for an early generation of $N$-body simulations that attempted to reproduce the phase-space properties of %Sgr 
the stream \citep{HelmiWhite2001sgr, Helmi2004sgr, Johnston2005sgr, Law2005sgr, Fellhauer2006sgr, Penarrubia2010sgr}. These works culminated in the landmark model of \citet{LM2010modelSgrStream}, which was capable of reproducing most of Sgr stream's features known at the time.

\defcitealias{Vasiliev2021tango}{V21}
Thanks to the \textit{Gaia} mission \citep{GaiaMission, gaiadr1, gaiadr2, GaiaEDR3Summary}, precise astrometric data for more than a billion stars in the Milky Way %, as well as in its satellite galaxies \citep{gaiaKinematics, GaiaEDR32021magClouds},
are now available, revolutionizing our views of the Milky Way and % This crucial piece of information, i.e., proper motions and parallaxes, has revolutionized
the knowledge about Galactic stellar streams \citep[e.g.,][]{Malhan2018streams, PriceWhelan2018gd1, Shipp2019streamsPMs, Riley2020streams, Li2021AtlasAliqaUma}. For instance, it has allowed the blind detection of $\mathcal{O}(10^5)$ high-probability members of Sgr stream \citep{Antoja2020SgrStream, Ibata2020SgrStream, Ramos2022sgr}% even in the absence of $v_{\rm los}$ values
, dramatically advancing our understanding of its present-day %dynamical status
kinematics. Moreover, a misalignment between the stream's track and the motion of its debris has been identified toward the leading arm (Galactic latitude $b > 0^{\circ}$) of Sgr \citep[][hereafter \citetalias{Vasiliev2021tango}]{Vasiliev2021tango}. Such observation can be reconciled with time-dependent perturbations induced by %a massive (${\geq}10^{11} M_\odot$) 
the Large Magellanic Cloud (LMC; see \citealt{Oria2022Sgr_bifurcations} and \citealt{Wang2022arXivSgr}).

\defcitealias{Penarrubia2021sgr}{PP21}
Despite these \textit{Gaia}-led advances, a fundamental difficulty in studies of Sgr stream continues to be the large heliocentric distances of its member stars (${\gtrsim}10$\,kpc as informed by, e.g., the aforementioned \citetalias{Vasiliev2021tango} model). %At such distances, stars are usually faint and, hence, the uncertainties in \textit{Gaia}'s measured parallaxes are exceedingly large and exposure times for spectroscopy become prohibitively long. 
This challenge is usually tackled via the utilization of stellar standard candles appropriate for the study of old stellar populations such as blue horizontal-branch and RR Lyrae stars \citep[e.g.,][]{Belokurov2014sgr, Hernitschek2017sgr, Ramos2020SgrStream}, allowing us to identify Sgr stream in angular-momentum space \citep[][hereby \citetalias{Penarrubia2021sgr}]{Penarrubia2021sgr} or in integrals of motion \citep{Yang2019sgrLAMOST, Yuan2019cetus, Yuan2020lms1}. %For instance, \citet[][hereby \citetalias{Penarrubia2021sgr}]{Penarrubia2021sgr} have recently utilized a sample of blue horizontal-branch and red giant-branch stars %\footnote{This compilation was originally described in \citet{PetersenPenarrubia2021} with data from \citet{Xue2011bhb, Xue2014_RGBcatalog, Xue2015break}.} 
%to demonstrate that Sgr stream can be isolated from other halo substructures in angular-momentum space% with a mixture modelling approach
%. Others have used similar tracers to detect the signature of Sgr stream in integrals of motion  \citep{Yang2019sgrLAMOST, Yuan2019cetus, Yuan2020lms1}.%, but with the caveat that these quantities (e.g., orbital actions) are dependent on the assumed potential model
% usage of standard candles (RRL, BHB, RGB)
% sgr in angular momentum (penarrubia+2021)
% add detection in 'IoM' space by yuan2019, 2020

Although the usage of some specific halo tracers has been crucial for advancing our knowledge of the dynamical status of Sgr stream, it comes with the obvious caveat of limited sample sizes. %Also, restricting our analyzes to certain spectral types and/or evolutionary stages can introduce biases in our interpretations of entire stellar populations. 
One way to go about this is to leverage both spectroscopic and photometric information from large-scale surveys in order to obtain full spectro-photometric distance estimates \citep{Santiago2016starhorse, Coronado2018lamost, McMillan2018rave, Queiroz2018, Hogg2019specphot, Leung2019distances} for much larger stellar samples. %In combination with informative priors from \textit{Gaia}'s parallaxes, this approach allows us to break the dwarf--giant degeneracy, a strategy pedagogically exemplified by \citet{sestito2019}% for low-metallicity stars
% usage spectro-photo-geometric distances
% h3 survey (johnson+2020 & naidu+2020)
% starhorse + APOGEE (hayes+2020)

\defcitealias{Johnson2020sgr}{J20}
Recently, \citet{Hayes2020} used spectro-photometric distances %computed %, by \citet{Queiroz2020}, 
for stars observed during the Apache Point Observatory Galactic Evolution Experiment \citep[APOGEE;][]{apogee2017} to investigate %$\alpha$-element 
abundances in the Sgr system (stream$+$remnant). Significant chemical differences between the leading and trailing ($b < 0^{\circ}$) arms were reported, with the latter being more metal-rich (by ${\sim}0.3$\,dex) than the former (see also \citealt{Monaco2005sgr, Monaco2007sgr}, \citealt{JingLi2016sgr, JingLi2019sgr}, \citealt{Carlin2018sgr}, and \citealt{Ramos2022sgr}), as well as $\rm[Fe/H]$ and [$\alpha$/Fe] gradients \textit{along} the stream itself (e.g., \citealt{Bellazzini2006sgr}, \citealt{Chou2007sgr, Chou2010sgr}, \citealt{Keller2010sgr}, \citealt{Shi2012sgr}, and \citealt{Hyde2015sgr}). Moreover, \citet[][referred to as \citetalias{Johnson2020sgr}]{Johnson2020sgr} investigated the stellar population(s) of Sgr stream with data from the Hectochelle in the Halo at High Resolution (H3; \citealt{Conroy2019surveyH3}) survey and spectro-photometric distances derived as in \citet[][see also \citealt{naidu2020}]{Cargile2020minesweeper}. The extended metallicity range (reaching $\rm[Fe/H] \approx -3$) probed by H3, in comparison to APOGEE ($\rm[Fe/H] \gtrsim -2$; see \citealt{Limberg2022gse} for a discussion), allowed these authors to uncover a metal-poor, dynamically hot, and spatially diffuse component of Sgr stream, confirming a previous suggestion by \citet{Gibbons2017sgr}.
% spectroscopic surveys (APOGEE, SEGUE)
% stellar population gradientsx
% usage of standard candles

In this contribution, we explore the phase-space and chemical properties of Sgr stream, but focusing on its very metal-poor (VMP; $\rm[Fe/H] < -2$)\footnote{Following the convention of \citet{beers2005}.} population. %However, instead of dividing this substructure into distinct components (dynamically cold/metal-rich vs. dynamically hot/metal-poor) as in, e.g., \citet{Gibbons2017sgr} and \citetalias{Johnson2020sgr}, we 
seeking to quantify the whole evolution of its kinematics as a function of chemistry. For this task, we need a large enough sample of stars covering a wide metallicity range. %Hence, we do not rely on high-resolution ($\mathcal{R} \geq 20{,}000$) spectroscopy as was done by \citet[][%also \citealt{Hasselquist2019sgr, Hasselquist2021dwarf_gals}, \citealt{Horta2022haloSubs}, and \citealt{Limberg2022gse}
%]{Hayes2020} and \citetalias{Johnson2020sgr}, but rather turn our attention to low-resolution ($\mathcal{R} \sim 1800$) data from the Sloan Extension for Galactic Understanding and Exploration \citep[SEGUE;][]{yanny2009, Rockosi2022segue} survey. 
Therefore, our attention was drawn to low-resolution ($\mathcal{R} \sim 1800$) spectroscopic data from the Sloan Extension for Galactic Understanding and Exploration \citep[SEGUE;][]{yanny2009, Rockosi2022segue} survey, a sub-project within the Sloan Digital Sky Survey \citep[SDSS;][]{sdssYork}. 
Atmospheric parameters provided by the SEGUE Stellar Parameter Pipeline \citep[SSPP;][]{AllendePrieto2008sspp, lee2008a, lee2008b, Lee2011sspp, Smolinski2011sspp} are combined with \textit{Gaia}'s parallaxes and broad-band photometry from various sources, similar to \citet{Queiroz2020}, to estimate spectro-photometric distances for ${\sim}175{,}000$ low-metallicity ($-3.5 \lesssim \rm[Fe/H] \leq -0.5$) stars in the SEGUE catalog. The complete description of this effort, including other spectroscopic surveys, 
is reserved for an accompanying paper (Queiroz et al., submitted).

This work is organized as follows. Section \ref{sec:data} describes the observational data analyzed throughout this work% as well as provide details about the derivation of spectro-photometric distances
. Section \ref{sec:sgr_stream_segue} is dedicated to investigate the chemodynamical properties of Sgr stream in SEGUE. Comparisons with the %state-of-the-art
$N$-body model of \citetalias[][]{Vasiliev2021tango} are presented in Section \ref{subsec:model}. We explore $\alpha$-element and carbon abundances in Section \ref{sec:abundances}. Finally, Section \ref{sec:conclusions} is reserved for a brief discussion and our concluding remarks.

\section{Data} \label{sec:data}

\subsection{SEGUE\texorpdfstring{$+$}xGaia and \texorpdfstring{$\mathtt{StarHorse}$}x} \label{subsec:segue+gaia}

%The main data set employed in this work is from SEGUE, a sub-project within the Sloan Digital Sky Survey \citep[SDSS;][]{sdssYork}. The 
SEGUE's emphasis on the distant halo is suitable for studying Sgr \citep[and other streams;][]{Newberg2010orphan, Koposov2010gd1} and has, indeed, been extensively explored for this purpose \citep{Yanny2009sgr, Belokurov2014sgr, deBoer2014sgr, deBoer2015sgr, Gibbons2017sgr, Chandra2022echoes, Thomas&Battaglia2022cetus}. The novelty is the availability of complete phase-space information thanks to \textit{Gaia}% and newly obtained spectro-photometric distances for SEGUE targets% by Queiroz et al. (submitted)
. Hence, we are in a position to construct a larger sample of confident Sgr stream members than previous efforts% (Section \ref{subsec:selection})
. 

Stellar atmospheric parameters, namely effective temperatures ($T_{\rm eff}$), surface gravity ($\log g$), and metallicities (in the form of [Fe/H]), as well as $\alpha$-element abundances ([$\alpha$/Fe]) and $v_{\rm los}$ values for SEGUE stars were obtained via application of the SSPP\footnote{Over the years, the SSPP has also been expanded to deliver carbon \citep{carollo2012, lee2013, lee2017, lee2019, Arentsen2022cemp}, nitrogen \citep[][]{Kim2022nitrogen}, and sodium \citep{Koo2022sodiumSSPP} abundances (see Section \ref{sec:carbon}).} routines. The final run of the SSPP to SEGUE spectra was presented alongside the ninth data release (DR9) of SDSS \citep[][]{SDSS_DR9} and has been included, unchanged, in all subsequent DRs. Recently, \citet{Rockosi2022segue} reevaluated the internal precision of SSPP's atmospheric parameters for DR9, which are no worse than 80\,K, 0.35\,dex, and 0.25\,dex for $T_{\rm eff}$, $\log g$, and [Fe/H], respectively, across the entire metallicity and color ranges explored. Unless explicitly mentioned, we consider SEGUE's as our fiducial stellar parameters throughout the remainder of this paper. These also serve as input for the Bayesian isochrone-fitting code \texttt{StarHorse} \citep{Santiago2016starhorse, Queiroz2018}, which, in turn, provides ages and distances.

In this work, %For the SEGUE/\texttt{StarHorse} run% (Section \ref{subsec:starhorse})
we only consider spectra with moderate signal-to-noise ratio ($S/N > 20$ pixel$^{-1}$). We keep only those stars within $4500 < T_{\rm eff}/{\rm K} < 6500$, which is the optimal interval for the performance of SSPP. Moreover, we limit our sample to low-metallicity stars ($\rm[Fe/H] \leq -0.5$%; as in \citealt{Perottoni2022gse}
), which removes most of the contamination from the thin disk, but maintains the majority of Sgr stream members; out of 166 stars analyzed by \citet{Hayes2020}, only 5 (3\%) show $\rm[Fe/H] > -0.5$.
% typical errors and cleaning 

We cross-match the above-described SEGUE low-metallicity sample with \textit{Gaia}'s early DR3 (EDR3; \citealt{GaiaEDR3Summary}) using 1.5$''$ search radius%, which provides parallaxes and absolute proper motions%\footnote{During the preparation of this manuscript, the full DR3 from \textit{Gaia} has been made available \citep{GaiaDR32022arXiv}. Nevertheless, we reinforce that this new release does not contain new astrometric or photometric calibrations.}
. In order to ensure the good quality of the data at hand, we only retain those stars whose renormalized unit weight errors %of the reduced astrometric $\chi^2$ 
are within the recommended range ($\texttt{ruwe} \leq 1.4$; \citealt{Lindegren2020_AstromSol})%, see also \citealt{Fabricius2021})
. Parallax biases and error inflation are handled following \citet{Lindegren2020_PlxBias} and \citet[][]{Fabricius2021}, taking into account magnitudes, colors, and on-sky positions%\footnote{Code available at \url{https://gitlab.com/icc-ub/public/gaiadr3_zeropoint/-/tree/master/}; see \citet{Anders2022starhorseEDR3}.}
. Stars with largely negative parallax values ($\texttt{parallax} < -5$) are discarded. Also, we emphasize that %, in \texttt{StarHorse}, 
only those stars with an available parallax measurement are considered to ensure good distance results. Those stars with potentially spurious astrometric solutions are also removed ($\texttt{fidelity\_v2} < 0.5$; \citealt{Rybizki2022fidelity}).
% our match with Gaia and cleaning of the astrometric data

%% anna's suggestion figure (:
\begin{figure}[pt!]
\centering
\includegraphics[width=1.0\columnwidth]{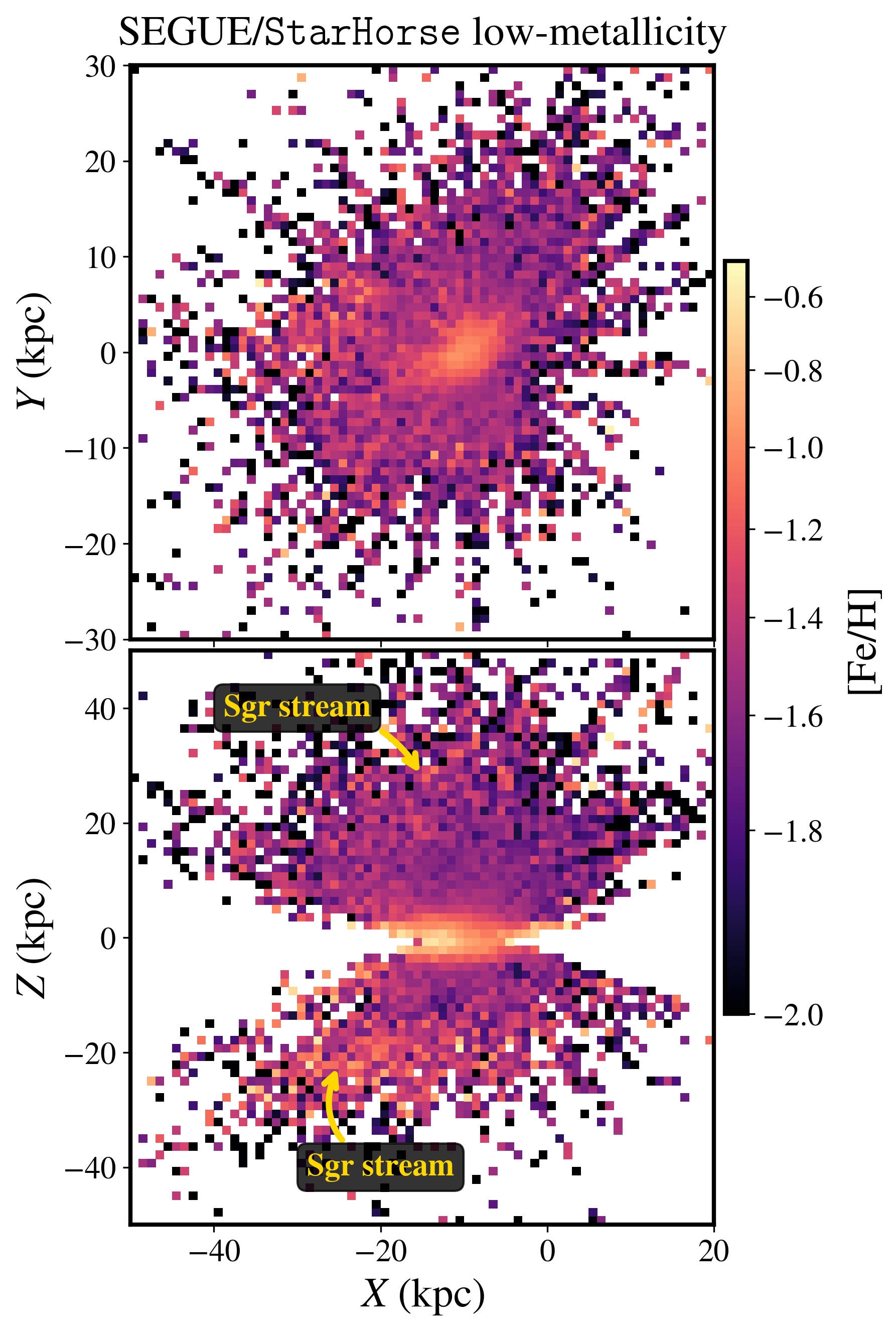}
\caption{Cartesian Galactocentric projections of the SEGUE/\texttt{StarHorse} low-metallicity sample. Top: $(X,Y)$. Bottom: $(X,Z)$. Spatial bins are color-coded by their mean [Fe/H] values. The attentive reader may notice the footprint of Sgr stream as metal-rich trails at ${|Z|} \gtrsim 20\,{\rm kpc}$%, for instance at $(X,Z) = (-30, -20)\,{\rm kpc}$ in the bottom panel
.
\label{fig:xyz}}
\end{figure}

%\subsection{The SEGUE/\texorpdfstring{$\mathtt{StarHorse}$}x Run} \label{subsec:starhorse}
% our new starhorse run and cleaning distances

We applied %the Bayesian isochrone-fitting code 
\texttt{StarHorse} to this SEGUE$+$\textit{Gaia} EDR3 sample in order to estimate precise distances that would allow us to study the Sgr stream; at ${\geq}10\,{\rm kpc}$ from the Sun, our derived %Gaussian 
uncertainties are at the level of ${\sim}13\%$.
%\texttt{StarHorse} \citep{Santiago2016starhorse, Queiroz2018} is an isochrone-fitting code capable of delivering stellar parameters, distances, and $V$-band ($\lambda = 542\,{\rm nm}$) extinctions for individual field stars in a Bayesian framework. We applied this method to the %metal-poor 
%SEGUE$+$\textit{Gaia} EDR3 sample %described in Section \ref{subsec:segue+gaia} %($\sim$175{,}000 stars) 
%in order to estimate precise distances that would allow us to study the Sgr stream; at ${\geq}10\,{\rm kpc}$ from the Sun, our derived %Gaussian 
%uncertainties are at the level of ${\sim}13\%$. 
The medians of the derived posterior distributions are adopted as our nominal values, while $16$th and $84$th percentiles are taken as uncertainties. %The initial mass function utilized is from \citet[][]{Chabrier2003imf}. 
Further details regarding stellar-evolution models and geometric priors can be found in \citet{Queiroz2018, Queiroz2020} and \citet{Anders2019starhorseGaiaDR2}. % A discussion on updated priors is presented in \citet[][]{Anders2022starhorseEDR3}. The full release of \texttt{StarHorse}'s data products for a variety of spectroscopic surveys% ($+$\textit{Gaia} EDR3)
%, including SEGUE for the complete metallicity range, will be made available alongside a parallel effort (Queiroz et al., submitted).
See also \citet{Anders2022starhorseEDR3} for details regarding the compiled photometric data.

%For this \texttt{StarHorse} run, spectroscopic (SEGUE) and astrometric (\textit{Gaia} EDR3) information were combined with multi-wavelength photometry from various large-area surveys. Specifically, the data employed came from the Two Micron All Sky Survey \citep[2MASS;][]{2MASS}, Wide-field Infrared Survey Explorer \citep[WISE;][in particular \citealt{Schlafly2019unWISE}]{WISEsurvey2010}, and Panoramic Survey Telescope and Rapid Response System \citep[Pan-STARRS1 DR1;][with corrections following \citealt{Scolnic2015panstarrs}]{Pan-STARRS2016}. Whenever available, $griz$ magnitudes from SkyMapper DR2 \citep[][with recalibrated zero points by \citealt{Huang2021recalibration}]{SkyMapperDR2}, are also included. SkyMapper's bluer bands ($u$ and $v$; \citealt{Bessell2011skymapper}) are discarded due to a limitation of the extinction law adopted \citep[][]{Schlafly2016extinction}. We refer the reader to \citet{Anders2022starhorseEDR3} for complete details regarding the compiled photometric data.

With the \texttt{StarHorse} output at hand, we restrict our sample to stars with moderate ($<$20\%) fractional Gaussian uncertainties in their estimated distance values. %Throughout the remainder of this paper, 
We refer to this catalog as the ``SEGUE/\texttt{StarHorse} low-metallicity sample'' or close variations of that. We refer the interested reader to \citet{Perottoni2022gse} for an initial %scientific 
application of these data. Its coverage in a Cartesian Galactocentric projection can be appreciated in Figure \ref{fig:xyz}. By color-coding this plot with the mean [Fe/H] in spatial bins, the footprint of Sgr stream is already perceptible as metal-rich trails at ${|Z|} \gtrsim 20\,{\rm kpc}$. %Finally, %The authors showed that the chemodynamical properties of stars in a pair of halo stellar overdensities \citep[][]{Vivas2001vod, Newberg2002ghosts, Belokurov2007HAC} are indistinguishable from nearby (within 5\,kpc from the Sun) debris of the \textit{Gaia}-Sausage/Enceladus \citep[GSE;][also \citealt{Haywood2018}]{belokurov2018, helmi2018} disrupted dwarf galaxy.

\subsection{Kinematics and Dynamics} \label{subsec:kindyn}

Positions ($\alpha$, $\delta$) and proper motions ($\mu_{\alpha}^{*} = \mu_{\alpha} \cos{\delta}$, $\mu_{\delta}$) on the sky, $v_{\rm los}$ values from SEGUE, and \texttt{StarHorse} heliocentric distances are converted to Galactocentric Cartesian phase-space coordinates using \texttt{Astropy} Python tools \citep{astropy, astropy2018}. %The Galactic fundamental parameters adopted are mostly the same as \citet{mcmillan2017}.
The adopted position of the Sun with respect to the Galactic center is $(X, Y, Z)_\odot = (-8.2, 0.0, 0.0208)\,{\rm kpc}$ \citep[][]{BlandHawthorn2016, Bennett&Bovy2019vertical}. The local circular velocity is $\mathbf{V_{\rm circ}} = (0.0, 232.8, 0.0)\,{\rm km\,s^{-1}}$ \citep{mcmillan2017}, while the Sun's peculiar motion with respect to $\mathbf{V_{\rm circ}}$ is $(U,V,W)_\odot = (11.10, 12.24, 7.25)\,{\rm km \ s^{-1}}$ \citep{schon2010}.
%distance from the Sun to the Galactic center is 8.2\,kpc \citep[][]{BlandHawthorn2016, EHT2022sgrA*I}. Within \texttt{Astropy}'s standard right-handed frame, $X_\odot = -8.2\,{\rm kpc}$. The local circular velocity is $\mathbf{V_{\rm circ}} = (0.0, 232.8, 0.0)\,{\rm km\,s^{-1}}$. The complete velocity vector of the Sun is $(V_x, V_y, V_z)_\odot = (11.10, 245.04, 7.25)\,{\rm km\,s^{-1}}$, which includes both $\mathbf{V_{\rm circ}}$ and the Sun's peculiar motion with respect to the local standard of rest being $(U,V,W)_\odot = (11.10, 12.24, 7.25)\,{\rm km \ s^{-1}}$ \citep{schon2010}. Lastly, the vertical displacement of the Sun with respect to the Galactic plane is $Z_\odot = 0.0208\,{\rm kpc}$ \citep{Bennett&Bovy2019vertical}.

%In Section \ref{subsec:selection}, the angular-momentum distribution of our sample will be utilized to select genuine members of the Sgr stream as suggested by previous works \citepalias[][]{Johnson2020sgr, Penarrubia2021sgr}. Hence, 
For reference, we write, below, how each component of the total angular momentum ($\boldsymbol{L}$) is calculated.

\begin{equation}
\begin{aligned}
L_x = YV_z - ZV_y \\
L_y = ZV_x - XV_z \\
L_z = XV_y - YV_x{,}
\end{aligned}
\end{equation}
where %($X, Y, Z, V_x, V_y, V_z$) corresponds to the complete phase-space vector in the Galactocentric Cartesian frame. Also, note that 
$L = \sqrt{L_x^2 + L_y^2 + L_z^2}$. We recall that, although $\boldsymbol{L}$ is not fully conserved in an axisymmetric potential, with the exception of the $L_z$ component, it has been historically used for the identification of substructure in the Galaxy as it preserves reasonable amount of clumping over time (see \citealt{Helmi2020} for a review)

For the entire SEGUE/\texttt{StarHorse} low-metallicity sample, we also compute other dynamical parameters, such as orbital energy ($E$) and actions ($\boldsymbol{J} = (J_R, J_\phi, J_z)$ in cylindrical frame). %In this work, 
%Actions are presented in the cylindrical frame, i.e., $\boldsymbol{J} = (J_R, J_\phi, J_z)$. %We recall that 
The azimuthal action is equivalent to the vertical component of angular momentum ($J_\phi \equiv L_z$) and we use these nomenclatures interchangeably. In order to obtain these quantities, orbits are integrated for 10\,Gyr forward with the \texttt{AGAMA} package \citep[][]{agama} within the axisymmetric Galactic model of \citet{mcmillan2017}. %The axisymmetric Galactic potential model of \citet{mcmillan2017} is adopted, which includes thin and thick stellar disks \citep{Gilmore1983thick}, gaseous disks \citep{DehnenBinney1998mass}, flattened bulge \citep{Bissantz2002bulge}, and spherical dark matter (DM) halo \citep{NFW1996halos}. Although the Milky Way contains non-axisymmetric features such as a central rotating bar and spiral arms (see \citealt{freeman2002}, \citealt{BlandHawthorn2016}, and \citealt{Barbuy2018} for reviews), these are not expected to significantly affect our calculations for halo stars. 
A total of 100 initial conditions were generated for each star with a Monte Carlo approach, accounting for uncertainties in proper motions, $v_{\rm los}$, and distance. The final orbital parameters are taken as the medians of the resulting distributions, with $16$th and $84$th percentiles as uncertainties.

\begin{figure*}[pt!]
\centering
\includegraphics[width=2.1\columnwidth]{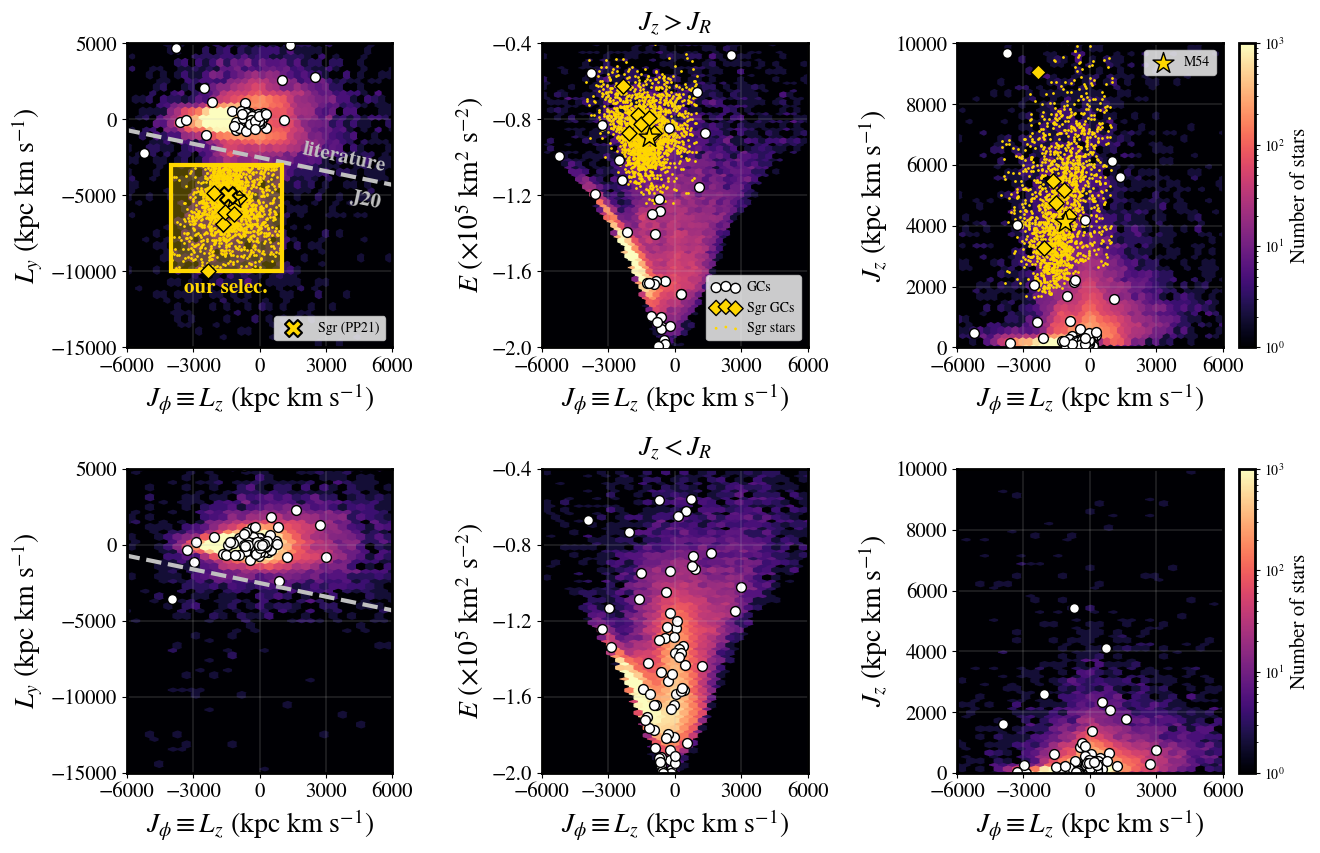}
\caption{Upper row: $J_z > J_R$ (predominantly polar orbits). Bottom: $J_z < J_R$ (radial/eccentric orbits). Background density maps are produced with the full SEGUE/\texttt{StarHorse} low-metallicity sample% (Section \ref{subsec:starhorse})
. White dots are Galactic GCs% compiled by \citet{VasilievBaumgardt2021gcs}
. Those associated with Sgr dSph/stream %(Section \ref{subsec:selection}) 
are displayed as yellow diamonds, with M54 as the star symbol (see text). Left panels: $(L_z, L_y)$. Our Sgr stream selection is shown as the yellow box. The gray dashed line exhibits the \citetalias{Johnson2020sgr} criterion. The yellow cross is the central location of Sgr in this space according to \citetalias{Penarrubia2021sgr}. Yellow dots represent %the kinematic/dynamical locus occupied by 
Sgr stream members. Middle: $(L_z, E)$. Right: $(L_z, J_z)$.
\label{fig:selection}}
\end{figure*}

\section{Sgr Stream in SEGUE} \label{sec:sgr_stream_segue}

\subsection{Selection of Members} \label{subsec:selection}

%In the context of 
Given our goals, %delineated by the end of Section \ref{sec:intro}, 
we seek to construct a sample of Sgr stream members that is both ($i$) larger in size and ($ii$) with greater purity than previously considered by \citetalias[][]{Johnson2020sgr}, but  ($iii$) with a similarly extended metallicity range, reaching the extremely metal-poor ($\rm[Fe/H] < -3$) regime. \citetalias[][]{Johnson2020sgr} have shown %, by comparison with the well-known \citet{LM2010modelSgrStream} model, 
that stars from the Sgr stream can be selected to exquisite completeness in the ($L_z, L_y$) plane, which exploits the polar nature of their orbits. %For the convenience of the reader, 
We reproduce their criterion in Figure \ref{fig:selection} (left panels, dashed lines). However, \citetalias[][]{Penarrubia2021sgr} have recently argued that the \citetalias{Johnson2020sgr}'s criterion also includes ${\approx}21\%$ of %contamination from Milky Way
interlopers. %Hence, we build on these previous efforts and design a new set of criteria capable of yielding better purity, but remaining comprehensible and readily reproducible.

We inspect the aforementioned ($L_z, L_y$) plane% (left panels of Figure \ref{fig:selection})
, splitting it into %Inspired by \citet[][their figure 6]{naidu2020}, we split this parameter space into 
$J_z > J_R$ (predominantly polar orbits; Figure \ref{fig:selection}, top row) and $J_z < J_R$ (radial/eccentric orbits; bottom row) as was done in \citet[][]{naidu2020}. %We notice that, 
In Figure \ref{fig:selection}, there exists an excess of stars toward negative values of $L_y$ (top left panel), which is prograde (top middle), and with $J_z \gtrsim 2000\,{\rm kpc\,km\,s^{-1}}$ \citep[top right;][]{Thomas&Battaglia2022cetus}, corresponding to the footprint of Sgr stream \citep[see][]{Malhan2022atlas}.
%. This same group of stars can be identified as a high-energy ($E \sim -0.8 \times 10^5\,{\rm km^2\,s^{-2}}$) prograde ($L_z \sim -1500\,{\rm kpc\,km\,s^{-1}}$) population (top middle panel); this is the footprint of Sgr stream \citep[see][]{Malhan2022atlas}. This substructure can also be recognized by its exceptional values of $J_z \gtrsim 2000\,{\rm kpc\,km\,s^{-1}}$ \citep[top right panel;][]{Thomas&Battaglia2022cetus}. %We also highlight that 
%This kinematic/dynamical signature of Sgr stream completely vanishes in the bottom panels. 
In the case where $J_z < J_R$, where Sgr stream completely vanishes, the ($L_z,E$) space is dominated by the \textit{ Gaia} Sausage/Enceladus \citep[GSE;][also \citealt{helmi2018}]{belokurov2018, Haywood2018}.%a prominent overdensity around $L_z \sim 0$ accompanied by %a multitude of 
%several globular clusters (GCs; white dots in Figure \ref{fig:selection}), which corresponds to\textit{ Gaia} Sausage/Enceladus \citep[GSE;][also \citealt{helmi2018}]{belokurov2018, Haywood2018}.

Developing on the above-described facts, we quantify how useful the $J_z > J_R$ condition is for eliminating potential GSE contaminants within our Sgr stream members. We look at a suite of chemodynamical simulations of Milky Way-mass galaxies with stellar halos produced by a single GSE-like merger presented in \citet[][]{Amarante2022gsehalos}. Within these models, the fraction of GSE debris that end up (at redshift $z=0$) on orbits with %whose 
$J_z > J_R$ is always below 9\%. Therefore, we incorporated this condition to our selection as it should remove ${>}90\%$ of potential GSE stars.

%We further inform ourselves with the \citetalias{Vasiliev2021tango} pure $N$-body model of the Sgr system. We verified that both star and dark matter (DM) particles are not expected to be found closer than ${\sim}6$\,kpc from the Sun. Hence, we eliminated stars with heliocentric distances within this range from our Sgr stream sample, mostly removing thin/thick-disk stars with low $J_z$ values.
%In practice, this cut removes mostly thin/thick-disk stars with low $J_z$ values.

\begin{figure*}[pt!]
\centering
\includegraphics[width=1.8\columnwidth]{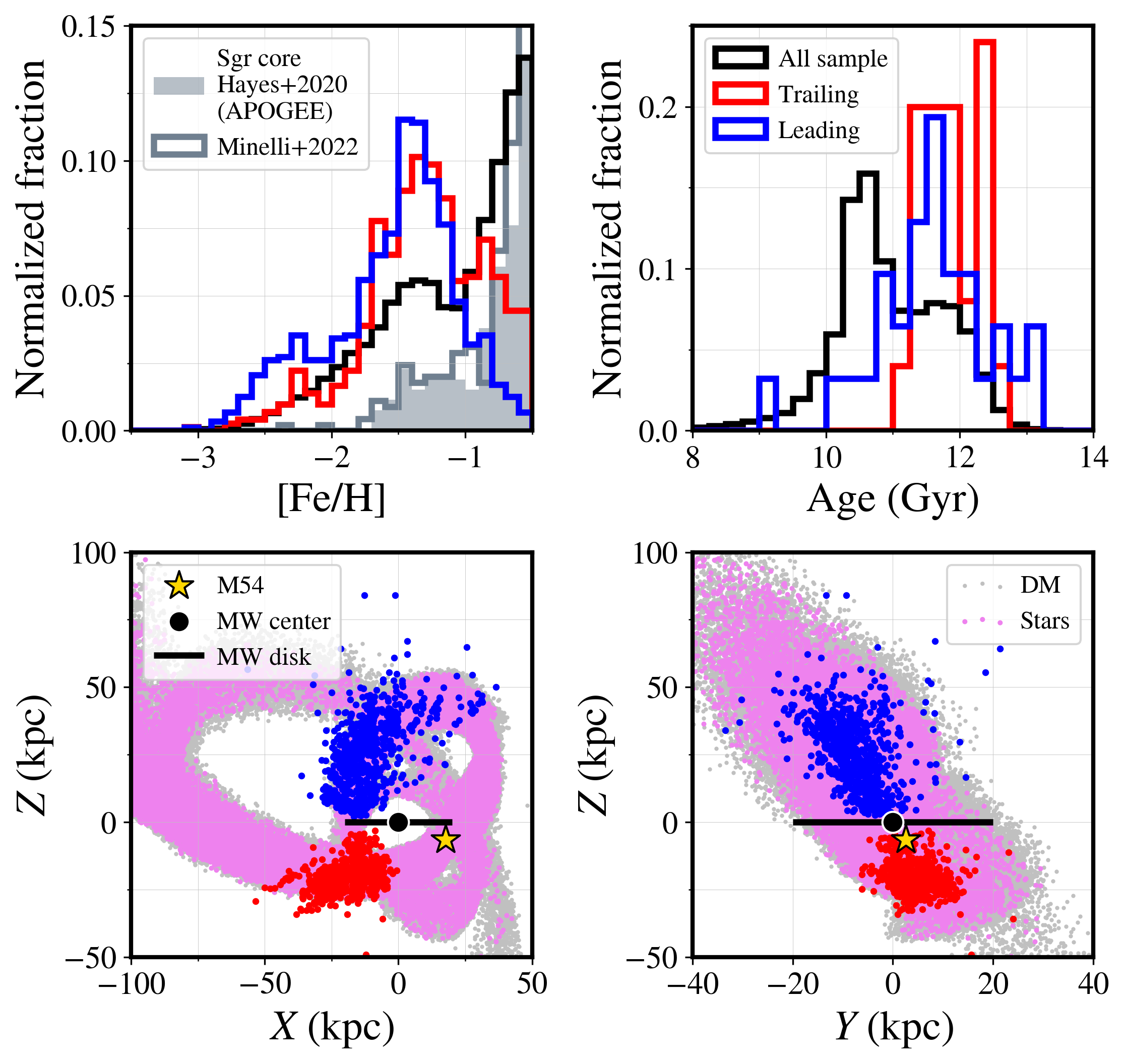}
\caption{Top row: metallicity (left) and age (right) distributions. In both panels, blue and red histograms represent the leading and trailing arms of Sgr stream, respectively% (Section \ref{subsec:selection})
. Filled and empty grayish histograms show the $\rm[Fe/H]$ distributions of Sgr's core according to \citet[][updated to APOGEE DR17]{Hayes2020} and \citet{Minelli2022SgrCoreMDF}, respectively. The complete SEGUE/\texttt{StarHorse} low-metallicity sample %(Sections \ref{subsec:segue+gaia} and \ref{subsec:starhorse})
is shown in black. %Age estimates are considered exclusively for turnoff and subgiant stars (Section \ref{subsec:arms}).
Bottom row: configuration space in the Galactic Cartesian coordinate system. Left: $(X,Z)$. Right: $(Y,Z)$. Blue and red dots are stars associated with the leading and trailing arms, respectively. The location of the M54 GC, which coincides with the center of Sgr dSph% \citep[e.g.,][]{Bellazzini2008m54}
, is shown as the yellow star symbol. The background Sgr model (stream$+$surviving core) is from \citetalias[][]{Vasiliev2021tango}, where gray and pink dots are dark matter and stellar particles, respectively. The Milky Way's center and disk (40\,kpc diameter) are illustrated by the black circle and line, respectively.
\label{fig:sgr_sample}}
\end{figure*}

Lastly, we restrict the kinematic locus occupied by Sgr stream in ($L_z, L_y$) in comparison to \citetalias{Johnson2020sgr} and consider only those stars at ${>}6$\,kpc from the Sun, in conformity with the \citetalias[][]{Vasiliev2021tango} model. %The results of \citetalias{Penarrubia2021sgr} (their figure 1) indicate that the Sgr stream footprint is better defined approximately within $-10 < L_y/(10^3 \,{\rm kpc} \,{\rm km} \,{\rm s}^{-1}) < -3$ and $-4 < L_z/(10^3 \,{\rm kpc} \,{\rm km} \,{\rm s}^{-1}) < +1$. %We note that the Galactic fundamental parameters \citep{Drimmel2018sunVel, GRAVITY2019} adopted by these authors are very similar to the ones described in our Section \ref{subsec:kindyn}. 
%An example of non-Sgr substructure that is also allowed by the \citetalias{Johnson2020sgr} criteria %, even after our $J_z > J_R$ cut, 
%is the Orphan stream \citep{Belokurov2007Orphan, Newberg2010orphan}, though this contamination appears to be minimal within the footprint of H3 \citep{Naidu2022mzr}. In Appendix \ref{sec:polar_streams}, we discuss the existence of other polar stellar streams \citep[see][]{Malhan2021lms1} that could overlap with Sgr in kinematic/dynamical parameter spaces. Overall, our set of criteria is %appears to be 
%robust against these potential sources of contamination.
In this work, the conditions that a star %within the SEGUE/\texttt{StarHorse} metal-poor sample 
must fulfill in order to be considered a genuine member of Sgr stream are listed below:
\begin{itemize}
    \vspace{-1mm}
    \item $J_z > J_R$;
    
    \vspace{-2mm}
    \item ${\rm heliocentric \ distance} > 6\,{\rm kpc}$;
    
    \vspace{-2mm}
    \item $-10 < L_y/(10^3 \,{\rm kpc} \,{\rm km} \,{\rm s}^{-1}) < -3$;
    
    \vspace{-2mm}
    \item $-4 < L_z/(10^3 \,{\rm kpc} \,{\rm km} \,{\rm s}^{-1}) < +1$.
\end{itemize}
This selection %of the Sgr stream 
is delineated by the yellow box in the top left panel of Figure \ref{fig:selection}. It is clear that our criteria is more conservative than \citetalias{Johnson2020sgr}'s. Nevertheless, the raw size of our final Sgr stream sample (${\sim}1600$ stars) is twice as large as the one presented by these authors %(${\sim}800$ members) 
despite the sharp cut at $\rm[Fe/H] < -0.5$. Moreover, the metallicity range covered %comfortably
reaches $\rm[Fe/H] \sim -3$, with ${\gtrsim}200$ VMP stars in the sample (top left in Figure \ref{fig:sgr_sample}). %This excess of VMP stars found in SEGUE is particularly suitable for us to quantify the %phase-space properties 
%kinematics of the diffuse, dynamically hot component of Sgr stream proposed by \citetalias[][]{Johnson2020sgr}. %Also, in Section \ref{sec:abundances}, we discuss how the identified VMP targets can help us trace the earliest chemical enrichment experienced by the Sgr system. 
Finally, although these Sgr stream candidates were identified from their locus in ($L_z, L_y$), we found them to be spatially cohesive and in agreement with the %spatial extent of the
\citetalias{Vasiliev2021tango} model in configuration space (bottom row of Figure \ref{fig:sgr_sample}). Even so, the potential contamination by other known polar streams \citep[see][]{Malhan2021lms1} is explored in Appendix \ref{sec:polar_streams}.

With this new set of selection criteria at hand, we verified which known Galactic globular clusters (GCs) would be connected to the Sgr system. Consequently, we examined the orbital properties %, computed as in Section \ref{subsec:kindyn}, 
of 170 GCs from the \textit{Gaia} EDR3-based catalog of \citet{VasilievBaumgardt2021gcs}. We found that a total of seven GCs can be linked to this group, including NGC 6715/M54, Whiting1, Koposov1, Terzan7, Arp2, Terzan8, and Pal12. We note that M54 has long been recognized to be the nuclear star cluster of Sgr dSph \citep[e.g.,][]{Bellazzini2008m54}. %We recommend \citet{Neumayer2020reviewNSC} for a recent review on the topic. 
Furthermore, most of these other GCs had already been attributed to Sgr by several authors \citep{massari2019, Bellazzini2020sgr, Forbes2020, Kruijssen2020kraken, Callingham2022gcs, Malhan2022atlas}. %Notorious absences in this list are NGC 2419 (marginally outside our selection), NGC 4147 and NGC 5634 (potential members of GSE\footnote{NGC 4147 and NGC 5634 have been associated with Helmi stream \citep{helmi1999} by, e.g., \citet{Callingham2022gcs}.}; \citealt{Limberg2022gse}), and NGC 5824 (recently associated with the Cetus\footnote{The Cetus stream was first described by \citet{Newberg2009cetus}.} accretion event \citep{Yuan2019cetus, Yuan2022cetus, Chang2020cetus, Malhan2022atlas}.

\subsection{Leading and Trailing Arms} \label{subsec:arms}
% talk about [Fe/H] differences between leading and trailing arms
% discuss how sgr stream appears to be older than thick-disk 

We begin our study of Sgr stream's stellar populations by looking at the metallicity distributions obtained for the leading and trailing arms and differences between them. In Figure \ref{fig:sgr_sample}, the immediately perceptible feature is the excess of VMP stars in the leading arm. On the contrary, the trailing arm presents a significant contribution of metal-rich ($\rm[Fe/H] \gtrsim -1$) stars. This property %of Sgr stream, the leading arm being more metal-poor than the trailing one, 
had already been noticed by several authors \citep[e.g.,][]{Carlin2018sgr, Hayes2020} and is recovered despite the intentional bias of the SEGUE catalog to low-metallicity stars (note the excess at $\rm[Fe/H] \lesssim -1$ in the black/all-sample histogram; see \citealt{Bonifacio2021mdf} and \citealt{Whitten2021splus} for discussions). 

The final median [Fe/H] values we obtained for the leading and trailing arms are $-1.46^{+0.02}_{-0.03}$ and $-1.28^{+0.03}_{-0.05}$, respectively, where upper and lower limits represent bootstrapped ($10^4$ times) 95\% confidence intervals. These metallicity values derived from SEGUE are $\sim$0.3--0.4\,dex lower than the ones obtained from APOGEE data \citep{Hayes2020, Limberg2022gse}, but we recall that this is due to SEGUE's target selection function \citep{Rockosi2022segue}.

We further notice that the location of these metallicity peaks for both arms of the stream are well aligned with the secondary, more metal-poor, $\rm[Fe/H]$ peak for stars in the core of Sgr (grayish histograms in Figure \ref{fig:sgr_sample}; \citealt{Hayes2020}\footnote{Updated with APOGEE DR17 \citep{APOGEEdr17}.}/APOGEE and \citealt{Minelli2022SgrCoreMDF}). Although differences in metallicity scales might be at play, this observation could be relevant for the evolution of the Sgr system in the presence of the Milky Way and LMC.

From \texttt{StarHorse}'s output, we should also, in principle, be able to access information regarding ages for individual stars as this parameter is a byproduct of the isochrone-fitting procedure (e.g., \citealt{Edvardsson1993isoFit}, \citealt{Jorgensen2005isoFit}, and \citealt{Sanders&Das2018ages}). However, there are some caveats in this approach. First, it becomes increasingly difficult to distinguish between isochrones of different ages toward both the cooler regions of the main sequence as well as the upper portions of the red giant branch (see figure 2 of  \citealt{souza2020} for a didactic visualization% of this phenomenon
). However, it is still possible to go around this issue by looking at the turnoff and subgiant areas where isochrones tend to be better segregated \citep[see discussion in][]{Vickers2021agesLAMOST}. Second, even at these evolutionary stages, variations in ages and metallicities have similar effects on the color--magnitude diagram (e.g., \citealt{YaleYonsei2001}, \citealt{YaleYonsei2004}, \citealt{Pietrinferni2004basti1, Pietrinferni2006basti2}, and \citealt{dotter2008}). Hence, spectroscopic [Fe/H] values can be leveraged as informative priors to break this age--metallicity degeneracy. Third, distant non-giant stars are quite faint, which is the case for our Sgr stream sample. This is where SEGUE's exquisite depth, with targets as faint as $g = 19.5$, where $g$ is SDSS broad band centered at $4800$\,{\AA} \citep{Fukugita1996sdssPhoto}, comes in handy.

%In this spirit, we selected stars in the SEGUE/\texttt{StarHorse} low-metallicity sample near the turnoff and subgiant branch regions in a first attempt to estimate the typical ages of both arms of Sgr stream. For the sake of consistency, instead of utilizing SEGUE's stellar parameters directly, we utilized those derived from \texttt{StarHorse} during the isochrone-fitting itself as these will be directly correlated with the ages at hand. Despite that, we note that .

In this spirit, we attempt to provide a first %, rough 
estimate of the typical ages for stars in Sgr stream. Similar to recent efforts \citep[][]{Bonaca2020, Buder2022halo, Xiang&Rix2022ages}, we selected stars in the SEGUE/\texttt{StarHorse} low-metallicity sample near the turnoff and subgiant stages. For the sake of consistency, for this task, we utilized stellar parameters derived by \texttt{StarHorse} itself during the isochrone-fitting process as these will be directly correlated with the ages at hand. These turnoff and subgiant stars are mostly contained within $4.5 < \log g_{\rm SH} \lesssim 3.6$ and $T_{\rm eff, SH} \gtrsim 5250\,{\rm K}$, where the subscript ``SH'' indicates values from \texttt{StarHorse} instead of SSPP. A parallel paper describes in detail this (sub)sample with reliable ages (Queiroz et al., submitted). %, including validations via comparisons with external catalogs. 
In any case, for the purpose of this work, we highlight that typical differences between SEGUE's atmospheric parameters and those obtained with \texttt{StarHorse} are at the level of SSPP's internal precision% (Section \ref{subsec:segue+gaia})
.

%For now, , but we note that, because SEGUE values serve as input for, differences between SSPP %atmospheric SEGUE and \texttt{StarHorse}
%parameters are at the level of the pipeline's internal precision (Section \ref{subsec:segue+gaia}).

We found a total of 56 turnoff or subgiant stars in Sgr stream (31 in the leading arm plus 25 in the trailing one) %with converged \texttt{StarHorse} solutions 
for which ages %we expect ages to be
are most reliable (top right panel of Figure \ref{fig:sgr_sample}). As expected, these are quite faint ($17.5 < g < 19.5$), which reinforces the value of a deep spectroscopic survey such as SEGUE. %The age distributions can be found in the top right panel of Figure \ref{fig:sgr_sample}. Overall, %the most striking feature is that 
Members of Sgr stream (blue and red histograms representing leading and trailing arms, respectively) appear to be older (11--12\,Gyr) than the bulk of %stars in 
our sample, which %we recall 
is mostly composed of thick-disk stars. It is reassuring that the age distribution for the entire SEGUE/\texttt{StarHorse} low-metallicity sample (black) peaks at 10--11\,Gyr, which is, indeed, in agreement with ages derived from asteroseismic data for the chemically-defined, i.e., high-$\alpha$, thick-disk population \citep[][]{SilvaAguirre2018age_disks, Miglio2021age_disks}. We quantify this visual interpretation with a kinematically-selected thick-disk sample, following $100 < |\mathbf{V}-\mathbf{V_{\rm circ}|/({\rm km\,s^{-1}}}) <180$ \citep[check][]{Venn2004, Bensby2014, LiZhao2017, Posti2018, Koppelman2020RESSONANCES}, where $\mathbf{V} = (V_x, V_y, V_z)$ is the total velocity vector of a given star, i.e., $V = \sqrt{V_x^2 + V_y^2 + V_z^2}$. Within the SEGUE/\texttt{StarHorse} low-metallicity data (${\sim}7,800$ stars), we found a median age of 10.6\,Gyr for this population.

For Sgr stream specifically, the bootstrapped median age for the leading arm is $11.6^{+0.4}_{-0.2}\,{\rm Gyr}$. For the trailing arm, we found $11.8^{+0.3}_{-0.2}\,{\rm Gyr}$. This translates to $11.7^{+0.3}_{-0.2}\,{\rm Gyr}$ considering all Sgr stream stars. Of course, uncertainties for individual stars are still substantial, usually at the level of ${>}25\%$ (${\sim}3$\,Gyr). Therefore, we hope that it will be possible to test this %tentative
scenario, that the Sgr stream is dominated by stars older (by ${\sim}1$\,Gyr) than those from the Galactic thick disk, with data provided by the upcoming generation of spectroscopic surveys, such as 4MOST \citep{4MOST2019}, SDSS-V \citep{Kollmeier2017}, and WEAVE \citep{WEAVE2016}, and building on the statistical isochrone-fitting framework of \texttt{StarHorse}.

\subsection{Evolution of Velocity Dispersion with Metallicity} \label{subsec:vel_disp}

% basically I want to say that the vel_disp increases with diminishing [Fe/H]
%As explicitly
%mentioned in Section \ref{sec:intro}, 
The original motivation for us to identify %SEGUE/\texttt{StarHorse}
Sgr stream members in the SEGUE/\texttt{StarHorse} catalog was to analyze the evolution of its kinematics extending deeply into the VMP regime%. Past efforts that conducted similar exercises include 
, similar to \citet[][]{Gibbons2017sgr} and \citetalias{Johnson2020sgr}. The former was the first to propose the existence of two populations in Sgr stream% using SEGUE data itself
. Its main limitation was the lack of complete phase-space information, which are now available thanks to \textit{Gaia}. Regarding the latter, the caveats were the small amount of (${\sim}50$) VMP stars %available 
in their %Sgr stream 
sample (from H3 survey%; see also \citealt{naidu2020}
) and potential contamination by Milky Way foreground stars \citepalias[][%see Section \ref{subsec:selection}
]{Penarrubia2021sgr}. Here, instead of splitting Sgr stream into two components, our approach is to model its velocity distribution %(e.g., \citealt{Li2017eridanusII, Li2018tucanaIII}) %as a whole 
across different [Fe/H] intervals. The results of this exercise can provide constraints to future chemodynamical simulations attempting to reproduce the Sgr system as was recently done for GSE \citep[][]{Amarante2022gsehalos}.

\begin{figure*}[pt!]
\centering
\includegraphics[width=2.\columnwidth]{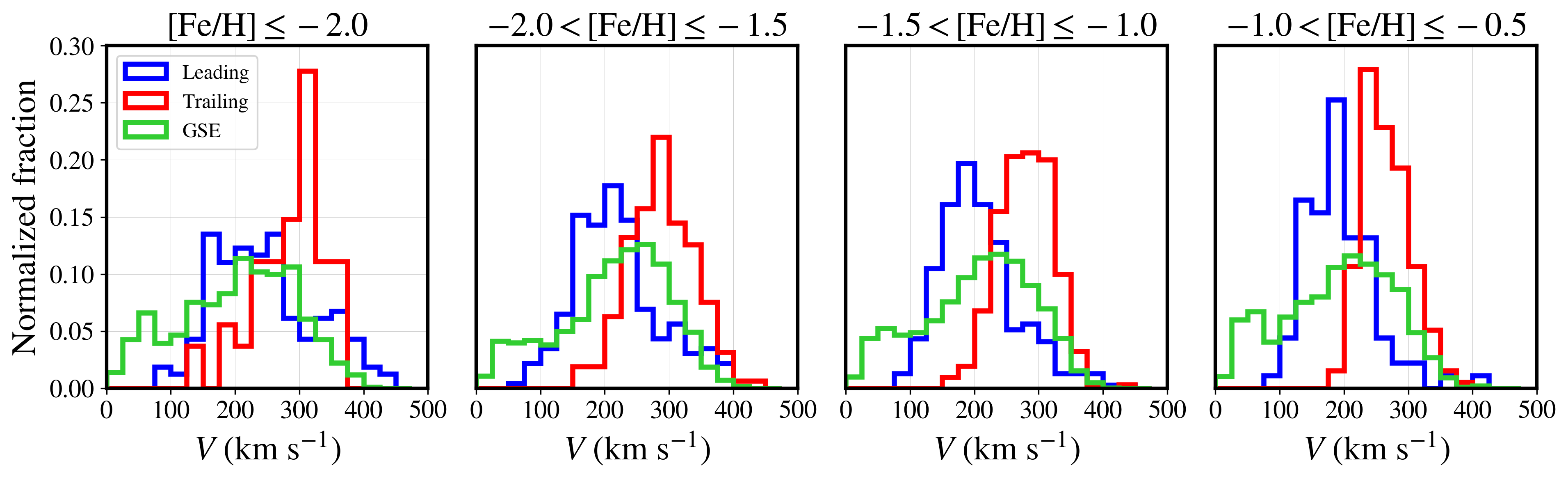}
\caption{Distributions of $V = \sqrt{V_x^2 + V_y^2 + V_z^2}$ in intervals of [Fe/H]. From left to right, we move from the VMP to the metal-rich regime. Blue, red, and green histograms represent the leading arm, trailing arm, and GSE, respectively% (Section \ref{subsec:vel_disp})
.
\label{fig:vel_disp}}
\end{figure*}

%In the context of the above-mentioned goal, 
Figure \ref{fig:vel_disp} displays the distributions of total velocity ($V$%; Section \ref{subsec:arms}
) across different metallicity ranges, from VMP (left) to metal-rich (%$\rm[Fe/H] > -1.0$; 
right). The color scheme is blue/red for leading/trailing arm as in Figure \ref{fig:sgr_sample}. From visual inspection, one can notice that both histograms become broader at lower [Fe/H] values. In order to %robustly 
quantify this effect of increasing velocity dispersion ($\sigma_V$) with decreasing [Fe/H], we model these distributions, while also accounting for uncertainties, using a Markov chain Monte Carlo (MCMC) method implemented with the \texttt{emcee} Python package \citep[][]{Foreman-Mackey2013emcee}. As in \citet[][]{Li2017eridanusII, Li2018tucanaIII}, the Gaussian $\log$-likelihood function is written as
\begin{equation}
\log{\mathcal{L}} = -\dfrac{1}{2} \sum_{i=1}^{N} \left[ \log{\left( \sigma^2_V + \sigma^2_{V,i} \right)} + \dfrac{  \left( V_i - \langle V \rangle \right)^2 }{ \left( \sigma^2_V + \sigma^2_{V,i} \right)} \right]{,}
\label{eq:likelihood}
\end{equation}
where $V_i$ and $\sigma_{V,i}$ are the total velocity and its respective uncertainty for the $i$th star within a given [Fe/H] bin% among those in Figure \ref{fig:vel_disp}
. We adopt only the following uniform priors: $0 < \langle V \rangle /({\rm km}\,{\rm s}^{-1}) < 500$ and $\sigma_V > 0$. Lastly, we run the MCMC sampler for 500 steps with 50 walkers, including a burn-in stage of 100. Although some of the $V$ histograms in Figure \ref{fig:vel_disp} show non-Gaussian tails, this exercise is sufficient for the present purpose.

The results of our MCMC calculations %with the %described MCMC strategy 
are presented in Table \ref{tab:vel_metal}. Upper and lower limits are 16th and 84th percentiles, respectively, from the posterior distributions. %Within the most metal-rich intervals, 
Between $-1.5 < \rm[Fe/H] \leq -0.5$, we found 
no statistically-significant (${<}1\sigma$%\footnote{In this work, whenever the Greek letter ``$\sigma$'' is utilized without any subscripts, it refers to Gaussian errors.}
) evidence for $\sigma_V$ variations. However, at $\rm[Fe/H] \leq -1.5$, the $\sigma_V$ %starts to 
increases substantially for both %leading and trailing 
arms. %Overall, we verified that, 
According to present data, the VMP component of Sgr stream (left panel of Figure \ref{fig:vel_disp}) is dynamically hotter than its metal-rich counterpart at the ${\gtrsim}2\sigma$ level. We also verified that this effect is less prominent (${\sim}1\sigma$) for GSE (green histograms in Figure \ref{fig:vel_disp}; Table \ref{tab:vel_metal}) even with a not-so-pure (at least 18\% contamination; \citealt{Limberg2022gse}) selection \citep[][]{Feuillet2020}, which is to be expected given the advanced stage of phase-mixing of this substructure.

%\vspace{-5mm}
\renewcommand{\arraystretch}{1.0}
\setlength{\tabcolsep}{1.0em}
\begin{table*}[ht!]
\centering
\caption{Velocity Dispersion for the Leading and Trailing Arms of Sgr, as well as GSE, in bins of [Fe/H]
}
\label{tab:vel_metal}
\begin{tabular}{>{\normalsize}c >{\normalsize}c >{\normalsize}c >{\normalsize}c >{\normalsize}c}
\hline
\hline
Substructure & $\sigma_V$ & \multicolumn{1}{c}{$\sigma_V$} & $\sigma_V$ & $\sigma_V$ \\ %
& (km\,s$^{-1}$) & \multicolumn{1}{c}{(km\,s$^{-1}$)} & (km\,s$^{-1}$) & (km\,s$^{-1}$) \\ %
& $\rm[Fe/H] \leq -2.0$ & \multicolumn{1}{c}{$-2.0 < \rm[Fe/H] \leq -1.5$} & $-1.5 < \rm[Fe/H] \leq -1.0$  & $-1.0 < \rm[Fe/H] \leq -0.5$    \\[1mm]
\hline
Leading  & $70^{+4}_{-4}\%$ $\phantom{0}$(194) & $62^{+3}_{-3}\%$ $\phantom{0}$(284) & $51^{+2}_{-2}\%$ $\phantom{0}$(405) & $53^{+5}_{-4}\%$ $\phantom{00}$(94) \\
Trailing & $44^{+6}_{-5}\%$ $\phantom{00}$(62) & $38^{+3}_{-3}\%$ $\phantom{0}$(185) & $32^{+2}_{-5}\%$ $\phantom{0}$(319) & $28^{+2}_{-2}\%$ $\phantom{0}$(203) \\
GSE      & $87^{+2}_{-2}\%$ (1153) & $86^{+1}_{-1}\%$ (4314) & $83^{+1}_{-1}\%$ (8020) & $82^{+2}_{-2}\%$ (1322) \\
\hline
%\multicolumn{5}{l}{\textbf{Note.} Upper and lower limits are 16th and 84th percentiles, respectively, derived from the MCMC sampling (Section \ref{subsec:vel_disp}).}
\end{tabular}
\end{table*}

Now, we put our results in context with those %obtained by the H3 survey team \citepalias[][]{Johnson2020sgr}
in the literature. With the understanding that Sgr stream is comprised of two kinematically distinct populations (\citetalias[][]{Johnson2020sgr}), the increasing $\sigma_V$ as a function of decreasing metallicity can be interpreted as larger fractions of the ``diffuse'' \citepalias[][]{Johnson2020sgr} component contributing to the low-[Fe/H] (dynamically hotter) bins. On the contrary, the ``main'' component, which contains most of the stars 
%(Section \ref{subsec:comps}) 
of the substructure, is preferentially associated with the high-[Fe/H] (dynamically colder) intervals. 

\citetalias{Penarrubia2021sgr} recently argued that the broad velocity distribution for metal-poor stars in Sgr stream could be an artifact of Milky Way contamination in the \citetalias[][]{Johnson2020sgr} Sgr stream data. However, this effect is still clearly present in our $2{\times}$ larger sample with more rigorous selection criteria. %We also stress that the higher $\sigma_V$ at lower [Fe/H] does not immediately imply that older wraps of the stream are necessarily dynamically hot. 
%We stress that
To summarize, in the low-metallicity regime, there appears to be considerable contribution from \textit{both} ancient and recently formed wraps of the stream. On the other hand, at high metallicities ($\rm[Fe/H] > -1.0$), only the newest wrap is represented% (see Section \ref{subsec:comps})
. %This will be made clear in Section \ref{subsec:comps} as we compare our data with the \citetalias[][]{Vasiliev2021tango} $N$-body model.

\begin{figure*}[pt!]
\centering
\includegraphics[width=2.1\columnwidth]{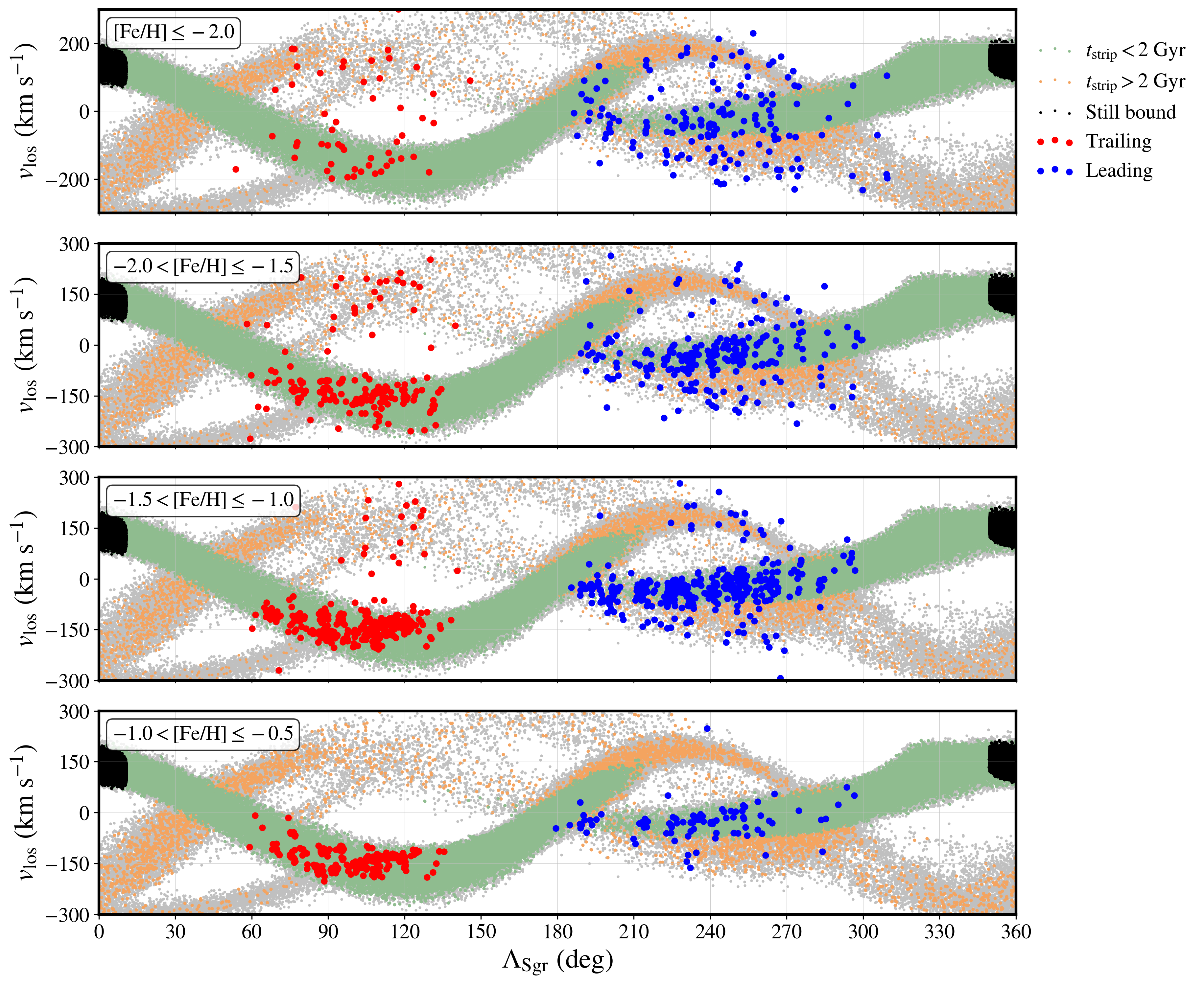}
\caption{Sgr stream in $(\Lambda_{\rm Sgr}, v_{\rm los})$ space. As in Figure \ref{fig:sgr_sample}, blue and red dots are stars associated with the leading and trailing arms, respectively% (Section \ref{subsec:arms})
. Dark matter and stellar particles from the \citetalias[][]{Vasiliev2021tango} model are shown as gray and colored dots, respectively. Green dots are attributed to the new ($t_{\rm strip}<2\,{\rm Gyr}$) and orange ones to the old ($t_{\rm strip}>2\,{\rm Gyr}$) wraps% (Section \ref{subsec:comps})
. %The new and old wraps of the stream are presented in green and orange, respectively
Black dots remain bound to the progenitor until the present day (redshift $z=0$), i.e., the end of the simulation. From top to bottom, we move from the VMP to the metal-rich regime following the same [Fe/H] ranges of Figure \ref{fig:vel_disp} and Table \ref{tab:vel_metal}.
\label{fig:sgr_coords}}
\end{figure*}

%\pagebreak
\section{Model Comparisons} \label{subsec:model}

In this section, we interpret the phase-space properties of Sgr stream and how they correlate with chemistry via the comparison of our Sgr sample with the \citetalias{Vasiliev2021tango} model. %Section \ref{subsec:V21model} summarizes the main features of the $N$-body model. Sections \ref{subsec:comps} and \ref{subsec:dsph} compare the observed properties of both trailing and leading arms against the simulation and how these are connected to the initial structure of Sgr, respectively.

\subsection{Model Properties, Assumptions, and Limitations} \label{subsec:V21model}

%Since we interpret our observations of the phase-space properties of Sgr stream and how they correlate with chemistry via the comparison of our Sgr sample with the model of \citetalias[][]{Vasiliev2021tango}, this section is dedicated to provide some information regarding these simulations' goals, constraints, and assumptions.
\citetalias[][]{Vasiliev2021tango}'s is a tailored $N$-body model of the Sgr system designed to match several properties of its tidal tails. In particular, in order to mimic the aforementioned misalignment between the stream track and its proper motions in the leading arm, the authors invoke the presence an LMC with a total mass of $1.5 \times 10^{11}\,M_\odot$, compatible with \citet{Erkal2019massLMC}, \citet{Shipp2021massLMC}, and \citet{Koposov2022LMCmass}. %This value is, indeed, in reasonable agreement with recent estimates of LMC's mass based on perturbations in Galactic stellar streams \citep{Erkal2019massLMC, Shipp2021massLMC, Koposov2022LMCmass}, but is slightly smaller than estimates based on the so-called ``timing argument'' \citep{Penarrubia2016timing} or cosmological abundance matching \citep{Boylan-Kolchin2010abundance, Dooley2017abundance}, both of which give ${\gtrsim}2\times10^{11}\,M_\odot$. %as well as require a flexible density profile for the Milky Way's halo (see their section 3.4), though the latter comes with the limitation that \citetalias[][]{Vasiliev2021tango} depend on the Chandrasekhar analytical prescription for dynamical friction \citep[e.g.,][]{MoBoschWhite2010book}%, an aspect recently criticized by \citet{Ramos2022sgr}
 The initial conditions are set to reproduce the present-day positions and velocities of both Sgr and LMC building on earlier results \citep{Vasiliev2020sgr}. However, unlike the LMC and Sgr, the Milky Way is not modeled in a live $N$-body scheme. Hence, it comes with the limitation that \citetalias[][]{Vasiliev2021tango} depend on the Chandrasekhar analytical prescription for dynamical friction \citep[e.g.,][]{MoBoschWhite2010book}. See \citet{Ramos2022sgr} for a discussion on how this approximation might influence the stripping history of the stream.

In the fiducial model, the initial stellar mass of Sgr dSph is $2\times10^8\,M_\odot$ and follows a spherical King density profile \citep{King1962}. Moreover, the system is embedded in an, also spherical, extended dark matter halo of $3.6 \times 10^9\,M_\odot$. Other key features of \citetalias{Vasiliev2021tango}'s work is the capability of recovering crucial kinematic and structural features of Sgr's remnant \citep[as in][]{Vasiliev2020sgr}, accounting for perturbations introduced by the gravitational field of LMC \citep{Garavito-Camargo2019lmc, Garavito-Camargo2021lmc, Cunningham2020lmc, Petersen2020reflex, PetersenPenarrubia2021, Erkal2021sloshing}, and properly following mass loss suffered by the system. 

%Of course, the work of \citetalias{Vasiliev2021tango} does not account for neither the presence of gaseous components nor star formation, hence supernova feedback, which have recently been shown to be relevant for the distribution of debris of disrupted dwarfs in the Galactic halo at redshift $z=0$ \citep{Wang2022arXivSgr}. 
Despite the close match between observations and the \citetalias{Vasiliev2021tango} model, there are a few limitations that could affect their results. For instance, the model does not account for the gaseous component %, thus no star formation
, which may be relevant for the distribution of the debris as discussed in \citet{Wang2022arXivSgr} and references therein. An additional caveat is the lack of bifurcations in the modeled stream, as originally observed by \citet{Belokurov2006Streams} and \citet[][see discussions by \citealt{Oria2022Sgr_bifurcations}]{Koposov2012sgr}. Finally, Sgr likely experienced at least one pericentric passage $\gtrsim$6\,Gyr ago as can be inferred from dynamical perturbations in the Galactic disk \citep[][and see \citealt{Antoja2018spirals}]{Binney2018spiral, Laporte2018sgr, Laporte2019sgr, Bland-Hawthorn2021spiral, McMillan2022spiral} as well as the star-formation histories of both the Milky Way \citep[][]{Ruiz-Lara2020sag} and Sgr itself \citep[][]{Siegel2007SFHsgr, deBoer2015sgr}. Hence, the \citetalias{Vasiliev2021tango} simulation, which starts only 3\,Gyr in the past, is unable to cover this earlier interaction.

\subsection{New and Old Wraps} \label{subsec:comps}

Figure \ref{fig:sgr_coords} shows observational and simulation data in $(\Lambda_{\rm Sgr}, v_{\rm los})$, where $\Lambda_{\rm Sgr}$ is the stream longitude coordinate as defined by \citet[][]{Majewski2003} based on Sgr's orbital plane. %This specific projection of phase space was chosen to illustrate our arguments simply because it can be readily compared with previous works \citepalias[e.g.,][]{Penarrubia2021sgr}. 
Leading/trailing arm stars are blue/red dots. These are overlaid to the \citetalias[][]{Vasiliev2021tango} model, where gray and colored points are dark matter and stellar particles, respectively. %As in \citetalias[][]{Penarrubia2021sgr}, 
We split these simulated particles according to their stripping time ($t_{\rm strip}$\footnote{Formally, $t_{\rm strip}$ is defined as the most recent time when a particle left a 5\,kpc-radius sphere around the progenitor \citepalias[][]{Vasiliev2021tango}.}). For the remainder of this paper, we refer to the portion of the (simulated) stream formed more recently ($t_{\rm strip} < 2\,{\rm Gyr}$) as the ``new'' wrap (green). The more ancient ($t_{\rm strip} > 2\,{\rm Gyr}$) component is henceforth the ``old'' wrap (orange). Stellar particles that are still bound to the progenitor by the end of the simulation (redshift $z=0$) are colored black.

%As was done %in our analysis of the evolution of $\sigma_V$ with [Fe/H]
%previously, 
We divide our Sgr stream data in the same metallicity intervals as Figure \ref{fig:vel_disp}/Table \ref{tab:vel_metal}. Essentially, our selected members of Sgr stream share all regions of phase space with the \citetalias[][]{Vasiliev2021tango} model particles. Notwithstanding, the bottom-most panel of Figure \ref{fig:sgr_coords} reveals a first interesting feature. Metal-rich %($\rm[Fe/H] > -1.0$)
stars in the sample are almost exclusively associated with the new wrap% ($t_{\rm strip} < 2\,{\rm Gyr}$, green)
, though this is more difficult to immediately assert for the leading arm because of the overlap between new and old %($t_{\rm strip} > 2\,{\rm Gyr}$, orange) 
portions within $180\degree \lesssim \Lambda_{\rm Sgr} < 300\degree$. %One way to, perhaps, confirm this apparent connection would be to kinematically decompose the stream as was done by \citetalias[][]{Johnson2020sgr} with a probabilistic approach, but that would also require the strong assumption that the diffuse/main components found by these authors can be directly associated with the new/old wraps seem in the \citetalias[][]{Vasiliev2021tango} model.
% overall, the data occupy the same regions of phase space as the model 
% metal-rich stars appear to be (almost) EXCLUSIVELY associated with the new wrap
% though this is difficult to assert for the leading arm as the new and ancient wraps overlap significantly in phase space

As we move toward lower-metallicity (upper) panels of Figure \ref{fig:sgr_coords}, we see larger fractions of observed Sgr stream stars coinciding with the old wrap in phase space. At the same time, the dense groups of stars overlapping with the new wrap fade away as we reach the VMP regime (top panel% of Figure \ref{fig:sgr_coords}
). As a direct consequence, stream members are more spread along the $v_{\rm los}$ axis in Figure \ref{fig:sgr_coords}, which, then, translates into the higher $\sigma_V$ discussed in Section \ref{subsec:vel_disp} for metal-poor/VMP stars. In general, the new wrap is preferentially associated with metal-rich %($\rm[Fe/H] > -1.0$)
stars, but also extends into the VMP realm. Conversely, the old component contains exclusively metal-poor ($\rm[Fe/H] \lesssim -1$) stars. Therefore, %the scenario that we envision is 
these suggest that, at low metallicities, Sgr stream is composed of a mixture between old and new wraps and this phenomenon drives the increasing $\sigma_V$ quantified in Table \ref{tab:vel_metal}.
% as we move toward the low-metallicity bins, we see larger fractions of stars overlapping with the ancient wrap
% as a consequence, we see larger spreads in vlos, hence showing the origin of the higher sigma_V at low [Fe/H] discussed in Section 3.3
% in general, the NEW WRAP contains more metal-rich stars, but extends to VMP
% on the other hand, the ANCIENT WRAP contains only metal-poor ([Fe/H] < -1) stars
% therefore, we envision that, at low metallicities, Sgr stream is composed of a mixture between old and new wraps. This phenomenon drives the increasing sigma_V

\begin{figure*}[pt!]
\centering
\includegraphics[width=2.1\columnwidth]{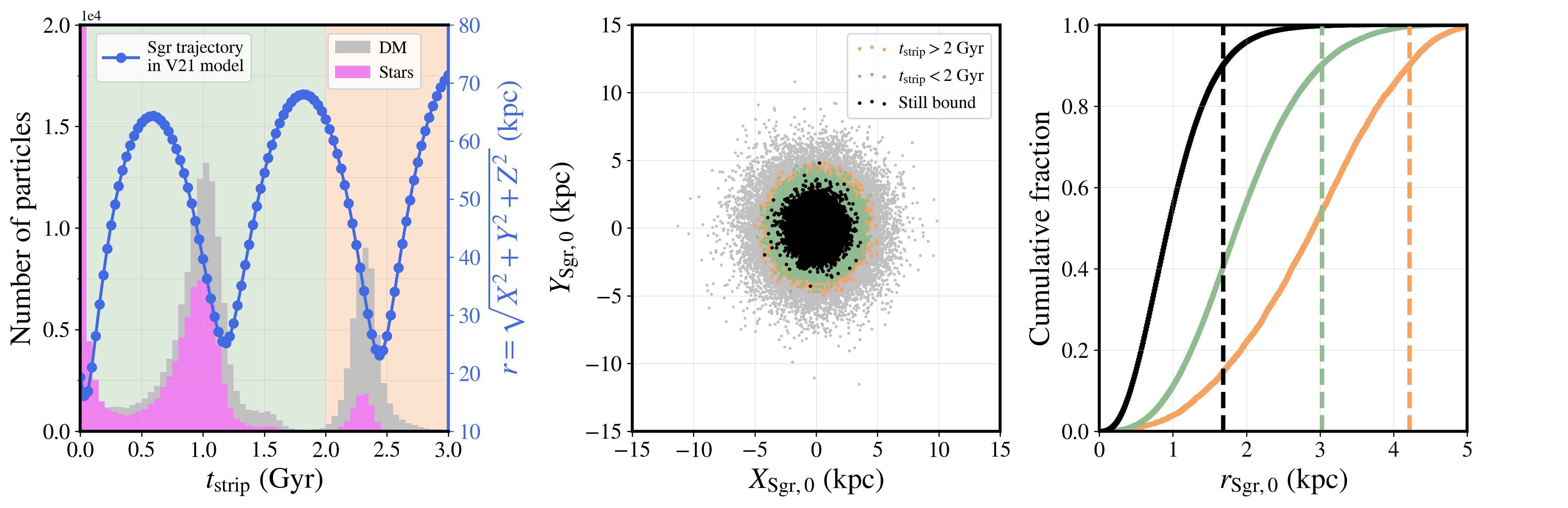}
\caption{Left: distributions of $t_{\rm strip}$ for dark matter (gray) and stellar (pink) particles in the \citetalias[][]{Vasiliev2021tango} model of Sgr (stream$+$dSph) system. The orbital trajectory of Sgr in the form of Galactocentric distance %($r = \sqrt{X^2 + Y^2 + Z^2}$) 
is presented as the overlapping blue line and circles. Middle: dark matter and stellar (colored dots) particles in configuration space, where $(X,Y)_{\rm Sgr,0}$ are spatial coordinates in a Sgr-centered sytem %Cartesian frame 
in the initial snapshot of the same \citetalias[][]{Vasiliev2021tango} model (see text). Right: cumulative distribution functions of galactocentric radii ($r_{\rm Sgr,0}$) of the same model particles, also centered around Sgr in the initial snapshot. Vertical dashed lines mark the positions containing 90\% of the stars of each component. In all panels, the color scheme is the same as Figure \ref{fig:sgr_coords}, where orange represents stars that end up as the old wrap of the stream ($t_{\rm strip} > 2$\,Gyr), green is for the new wrap ($t_{\rm strip} < 2$\,Gyr), and black are those particles that remain bound to Sgr dSph until present-day/end of the simulation ($t_{\rm strip} = 0$).
\label{fig:initial}}
\end{figure*}

%\pagebreak
\subsection{Sgr dSph Before its Disruption} \label{subsec:dsph}

The dichotomy between metal-rich/cold and metal-poor/hot portions of Sgr stream has been suggested, by \citetalias[][]{Johnson2020sgr}, to be linked to the existence of a stellar halo-like structure in Sgr dSph prior to its infall. This stellar halo would have larger velocity dispersion, be spatially more extended, and have lower metallicity than the rest of the Sgr galaxy. As a consequence of its kinematics, this component would be stripped at earlier times. Indeed, we verified that the old wrap of Sgr stream, stripped ${>}2\,{\rm Gyr}$ ago in \citetalias[][]{Vasiliev2021tango}'s model, is %solely
mainly associated with metal-poor stars (%Section \ref{subsec:comps}/
Figure \ref{fig:sgr_coords}), in conformity with \citetalias[][]{Johnson2020sgr}'s hypothesis. Meanwhile, the majority of the most metal-rich stars can be attributed to the new wrap. In order to check how the present-day properties of Sgr stream are connected to those of its dSph progenitor, hence testing other conjectures of \citetalias[][]{Johnson2020sgr}, we now look at the initial snapshot of \citetalias[][]{Vasiliev2021tango}'s simulation, including the satellite's %inferred
orbit and disruption history.
% J20 suggested that Sgr dSph had something like a 'stellar halo' before its infall
% this 'stellar halo' would be spatially more extended, dynamically hotter, and metal-poorer in comparison to the overall dwarf galaxy
% we can test this by looking at the INITIAL snapshot of the sim. (thx eugene)

%In order for us to comprehend the assembly of the stream over time according to the \citetalias[][]{Vasiliev2021tango} model, 
Simply to comprehend the assembly of the stream over time according to the \citetalias[][]{Vasiliev2021tango} model, we plot the distribution of $t_{\rm strip}$ in the left panel of Figure \ref{fig:initial}. %The color scheme is the same as Figure \ref{fig:sgr_sample}, with gray and pink representing dark matter and stars, respectively. 
The excess at $t_{\rm strip} = 0$ is due to $N$-body particles that remain bound to the progenitor. On top of these histograms, we add the trajectory of Sgr dSph in the simulation (blue line and dots) in terms of its Galactocentric distance% in the Cartesian system ($r = \sqrt{X^2 + Y^2 + Z^2}$; see additional axis)
. With this visualization, it is clear how intense episodes of material being stripped (at both $2.0 < t_{\rm strip}/{\rm Gyr} \lesssim 2.7$ and $0.5 \lesssim t_{\rm strip}/{\rm Gyr} \lesssim 1.5$) are intrinsically related to close encounters between Sgr and the Milky Way (at ${\sim}2.5$ and ${\sim}1.2$\,Gyr ago), which originates the new and old wraps discussed in Section \ref{subsec:comps}. Also, note how most of the material is associated with the recently formed component (new wrap) of the stream.
% first, we see that the formation of the new vs. ancient wraps is associated with different pericentric passages at ~1.2 and ~2.5 Gyr ago (this is where the rationale for splitting at 2 Gyr comes from)
% second, we can verify that most (~90%) of the stripped material is associated with the NEW WRAP, i.e., the most recent pericentric passage

In order to test \citetalias[][]{Johnson2020sgr}'s conjecture that the stripped portion of Sgr dSph associated with the formation of the old wrap was already dynamically hotter than the new one prior to the galaxy's disruption, we check the $\sigma_V$, with respect to Sgr, of these components in the initial snapshot of \citetalias[][]{Vasiliev2021tango}'s simulation, which starts 3\,Gyr in the past (redshift $z \sim 0.25$ in \citealt{PlanckCollab2020} cosmology). Indeed, the $\sigma_V$ of stars that end up forming the old wrap, i.e., stripped at earlier times, is higher (${\sim}18$\,km\,s$^{-1}$) in comparison to the $\sigma_V$ of stars from the new component (${\sim}14$\,km\,s$^{-1}$). %We also note that, in the initial snapshot, these dynamically hot and cold portions of Sgr appear to essentially overlap in configuration space as can be seem in middle panel of Figure \ref{fig:initial}. In this plot, $(X,Y)_{\rm Sgr,0}$ represent Sgr-centered Cartesian coordinates in the initial snapshot. 

In the middle panel of Figure \ref{fig:initial}, the initial snapshot is presented in configuration space as $(X,Y)_{\rm Sgr,0}$, a Sgr-centered frame. The orange dots ($t_{\rm strip} > 2$\,Gyr/old wrap) in this plot are less centrally concentrated (90\% of stellar particles within $\sim$4\,kpc) than the green ones ($t_{\rm strip} < 2$\,Gyr/new wrap; 90\% within 3\,kpc). This behavior is clear from the right panel of the same figure that shows the cumulative distributions of galactocentric radii ($r_{\rm Sgr,0}$) in the same system. Also, stars that remain bound until redshift $z=0$ have even lower $\sigma_V$ (${\sim}11$\,km\,s$^{-1}$) and are spatially more concentrated (90\% within ${<}2$\,kpc) than the other components.
% third, indeed, stars stripped earlier (>2Gyr, ancient wrap) were already dynamically hotter at the initial snapshot, but...
% (fourth) these are not necessarily in a configuration where they inhabit the outer regions/periphery of the galaxy. rather, they occupy the same region as those stripped at later times (<2Gyr, new wrap)

From the above-described properties of the \citetalias[][]{Vasiliev2021tango} model, we can infer that the periphery of the simulated Sgr dSph contains a larger fraction of stars that end up as the old wrap (stripped earlier) in comparison to its central regions. Therefore, with the understanding that the old wrap is essentially composed of low-metallicity stars), we reach the conclusion that the core regions of Sgr dSph were more metal-rich than its outskirts prior to its accretion.
% (fifth) nevertheless, stars stripped more recently are spatially more concentrated in comparison to those stripped earlier
% important to connect to the idea of metallicity gradient within the dwarf galaxy
We recall that, indeed, previous works reported evidence for a metallicity gradient in the Sgr remnant \citep[][]{Bellazzini1999sgr, Layden2000sgr, Siegel2007SFHsgr, McDonald2013sgr, Mucciarelli2017sgr, Vitali2022SgrPristine}. Nevertheless, fully understanding how these stellar-population variations in the Sgr system relate to %star-formation bursts induced by 
its interaction with the Milky Way remains to be seem (for example, via induced star-formation bursts; \citealt{Hasselquist2021dwarf_gals}).

Although our interpretation %based on the combination between observed data (from SEGUE/\texttt{StarHorse}) and the $N$-body model from \citetalias[][]{Vasiliev2021tango} 
favors a scenario where Sgr dSph %was able 
had enough time to develop a metallicity gradient before its disruption, quantifying this effect is difficult. %One way to approach this would be to kinematically decompose Sgr stream stars, as in \citetalias[][]{Johnson2020sgr}, then rearranging them into spatial distributions following the same density profiles of the different components (early or late stripping) of the \citetalias[][]{Vasiliev2021tango} model. Unfortunately, this strategy is difficult to be applied for the SEGUE data because of its selection function% that yields an excess of metal-poor stars and would require correction
%. The other way around is also feasible, i.e., 
One way to approach this would be \textit{painting} the model with \textit{ad hoc} metallicity gradients and, then, comparing with, for instance, the present-day [Fe/H] variations observed across the Sgr stream (\citealt{Hayes2020} and references therein). We defer this exploration to a forthcoming paper.

\section{Chemical Abundances} \label{sec:abundances}

\subsection{\texorpdfstring{$\alpha$}x Elements} \label{sec:alpha}

Apart from $T_{\rm eff}$, $\log g$, and $\rm[Fe/H]$, the SSPP also estimates $\alpha$-element abundances based on the wavelength range of $4500 \leq \lambda/{\rm \angstrom} \leq 5500$ \citep[][]{Lee2011sspp}, which contains several \ion{Ti}{1} and \ion{Ti}{2} lines as well as the \ion{Mg}{1} triplet (${\sim}5200$\,{\AA}).
%The strategy adopted in SSPP to estimate $\alpha$-element abundances is to match the observed spectra with synthetic ones within the wavelength range $4500 \leq \lambda/{\rm \angstrom} \leq 5500$ \citep[][]{Lee2011sspp}. This region contains several absorption features of interest, including several \ion{Ti}{1} and \ion{Ti}{2} lines as well as the \ion{Mg}{1} triplet (${\sim}5200$\,{\AA}), but also avoids the CH $G$-band at ${\sim}4300$\,{\AA}. Indeed, 
\citet[][]{deBoer2014sgr} utilized $[\alpha/{\rm Fe}]$ %\footnote{Formally, [$\alpha$/Fe] is weighted by line strengths; $[\alpha/{\rm Fe}] = 0.5\rm[Mg/Fe] + 0.3\rm[Ti/Fe] + 0.1\rm[Si/Fe] + 0.1\rm[Ca/Fe]$ \citep[][]{Lee2011sspp}.}
values made available by SEGUE for stars in Sgr stream to argue that a ``knee'' %\citep[e.g.,][]{Matteucci1990} 
existed at $\rm[Fe/H] \lesssim -1.3$ in the [$\alpha$/Fe]--[Fe/H] diagram \citep[][]{Wallerstein1962gdwarfs, Tinsley1979} for this substructure. However, this result is not supported by contemporaneous high-resolution spectroscopic data, specially from APOGEE \citep[][]{Hayes2020, Horta2022haloSubs, Limberg2022gse}, also H3 \citep[][]{Johnson2020sgr, naidu2020}. If the position of Sgr's $\alpha$ knee was truly located at such high [Fe/H], it would imply that it should be even more massive than GSE under standard chemical-evolution prescriptions \citep[e.g.,][]{Matteucci1990}. %\citep[e.g.,][]{Monty2020, Horta2022haloSubs}. 
%In fact, the [$\alpha$/Fe] vs. metallicity distribution of Sgr flattens at $\rm[Fe/H] \gtrsim -1$, which is a telltale sign that this dSph galaxy experienced additional burst(s) of star formation, likely due to its interaction with the Milky Way \citep[][]{Hasselquist2021dwarf_gals}.

\begin{figure*}[pt!]
\centering
\includegraphics[width=2.1\columnwidth]{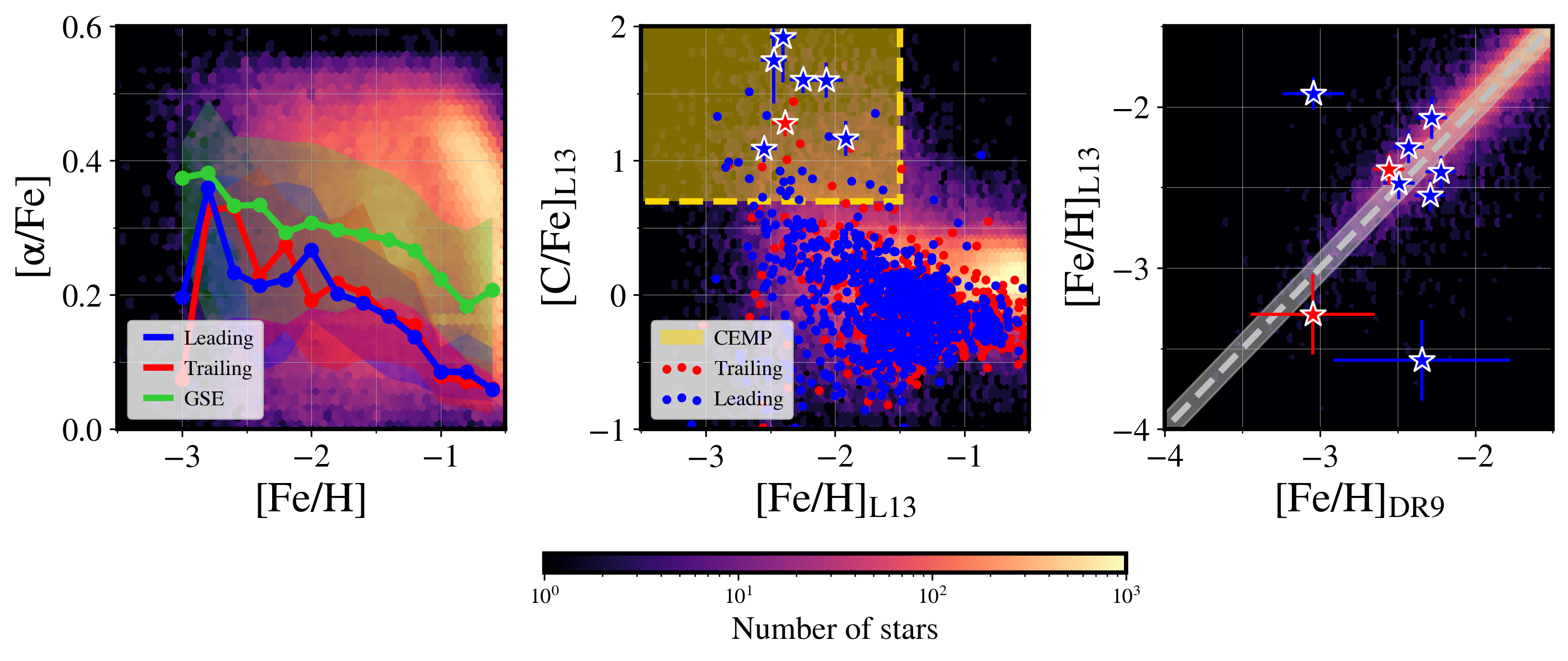}
\caption{Left: [$\alpha$/Fe]--[Fe/H]. Green line is the running median of GSE's [$\alpha$/Fe] values in bins of 0.2\,dex in [Fe/H], with the shaded area covering 16th and 84th percentiles. Blue and red lines and shaded regions are the same, but for the leading and trailing arms of Sgr stream, respectively. Middle: [C/Fe]--[Fe/H]. The yellow rectangle marks the locus of CEMP stars. Blue/red symbols are leading/trailing arm stars. Candidate ($3\sigma$; see text) CEMP stars are shown as star symbols. Right: $\rm[Fe/H]_{\rm L13}$--$\rm[Fe/H]_{\rm DR9}$, where the ``L13''and ``DR9'' subscripts refer to [Fe/H] values either from \citet[][]{lee2013} or SEGUE's standard catalog from SDSS DR9%(\citealt{SDSS_DR9}%/Section \ref{subsec:segue+gaia}
%)
. Background %two-dimensional 
density maps represent the full SEGUE/\texttt{StarHorse} low-metallicity sample% (Section \ref{subsec:starhorse})
.
\label{fig:chem}}
\end{figure*}

%We take the opportunity, in this work, to 
Here, we revisit the $\alpha$ abundances for Sgr stream using SEGUE, but with a larger sample with lower contamination% than previously considered
. In the left panel of Figure \ref{fig:chem}, we see the continuous decrease of [$\alpha$/Fe] as a function of increasing metallicity for both the Sgr stream and GSE%, which is expected for standard chemical-evolution prescriptions \citep[e.g.,][]{Matteucci1990}
. Most important, at a given value of [Fe/H], the median [$\alpha$/Fe] of Sgr stream (both leading and trailing arms) is lower than GSE's. This difference becomes more prominent at $\rm[Fe/H] \gtrsim -1.5$, in agreement with the aforementioned high-resolution spectroscopy results from both H3 %\citep[][]{naidu2020, Naidu2022mzr} 
and APOGEE% \citep[][]{Hasselquist2021dwarf_gals, Horta2022haloSubs, Limberg2022gse}
. Despite that, the low accuracy/precision of [$\alpha$/Fe] in SEGUE still makes it difficult to attribute stars to certain populations on an individual basis.

%Overall, similarly to the age distributions presented in %Section \ref{subsec:arms}/
%Figure \ref{fig:sgr_sample}, [$\alpha$/Fe] seems to be capable of revealing broad disparities between halo/Milky Way components. Indeed, SEGUE's [$\alpha$/Fe] abundance ratios were used to investigate the chemical thin--thick disk decomposition, as well as the so-called Splash or \textit{heated} disk \citep[][]{DiMatteo2019, AnBeers2020blueprintI, AnBeers2021blueprintII, belokurov2020, Amarante2020splash}, by several authors in the past \citep[][]{Lee2011disk, Bovy2012disk, IvezicBeersJuric2012, Liu2012disk, Han2020disk, Lee2022disk}. However, the low accuracy/precision of [$\alpha$/Fe] in SEGUE still makes it difficult to attribute stars to certain populations on an individual basis.

\subsection{Carbon} \label{sec:carbon}

With the SEGUE low-metallicity data at hand, we also explore carbon abundances. In particular, we are interested in finding carbon-enhanced metal-poor (CEMP; $\rm[C/Fe] > +0.7$ and $\rm[Fe/H] < -1$; see \citealt{beers2005}, \citealt{aoki2007}, and \citealt{placco2014Carbon}) stars in Sgr stream. The reasoning for that being the recent results by \citet[][also \citealt{Hansen2018sgr} and \citealt{Chiti2019sgr}]{Chiti2020sgr} where these authors found no CEMP star in their sample of Sgr dSph members within $-3.1 < \rm[Fe/H] \lesssim -1.5$. Moreover, we utilize observations of Sgr stream as a shortcut to check for potential differences in CEMP fractions between a dwarf galaxy and the Milky Way's stellar halo \citep[see][]{Venn2012, Kirby2015carbon, Salvadori2015, Chiti2018sculptor} in a homogeneous setting. Given that the CEMP phenomenon, specially at $\rm[Fe/H] \lesssim -2.5$, is connected to nucleosynthesis events associated with the first generations of stars, perhaps Population III \citep[e.g.,][]{Nomoto2013, yoon2016, Chiaki2017}, identifying such objects provide clues about the first chemical-enrichment processes that happened in a galaxy.

Throughout this section, we consider carbon abundances obtained for SEGUE spectra by \citet[][see also \citealt{carollo2012}, \citealt{lee2017, lee2019}, and \citealt{Arentsen2022cemp}]{lee2013}% in an independent run of the SSPP
. Inconveniently, this catalog also comes with slight variations of the stellar atmospheric parameters in comparison with the public SEGUE DR9 release%, which was adopted for the new %large-scale
%\texttt{StarHorse} run% (Section \ref{subsec:segue+gaia})
. Therefore, in order to confidently identify CEMP stars, we first select candidates using only \citet[][]{lee2013} [C/Fe] and [Fe/H] (subscripts ``L13'' in Figure \ref{fig:chem}). Then, we compare $\rm[Fe/H]_{\rm L13}$ with [Fe/H] values from our standard DR9 sample ($\rm[Fe/H]_{\rm DR9}$; right panel of Figure \ref{fig:chem}) to confirm their low-metallicity nature.

The middle panel of Figure \ref{fig:chem} ([C/Fe]--[Fe/H]) exhibits our selection of CEMP candidates (yellow box). %The yellow box delineates the considered CEMP locus in [C/Fe]--[Fe/H] space.
Note that we take only those stars at $\rm[Fe/H]_{\rm L13} < -1.5$, for consistency with the metallicity range covered by \citet[][]{Chiti2020sgr}. Expanding this boundary to $\rm[Fe/H]_{\rm L13} < -1$ would only include a couple of additional CEMP candidates. We discovered a total of 39 likely-CEMP stars (33 at $\rm[Fe/H]_{\rm L13} < -2$). With this sample at hand, we looked for those candidates confidently ($3\sigma$ in $\rm[C/Fe]_{\rm L13}$) encompassed by the CEMP criteria. We found 7 such objects, shown as star symbols in Figure \ref{fig:chem}. Although two of these CEMP stars have discrepant metallicity determinations ($\rm[Fe/H]_{\rm L13}$ vs. $\rm[Fe/H]_{\rm DR9}$; right panel of Figure \ref{fig:chem}), we can still confidently assert that there exist CEMP stars in Sgr stream.

A possible explanation for the lack of CEMP stars in \citeauthor{Chiti2020sgr}'s (\citeyear{Chiti2020sgr}) sample could be their photometric target selection, which was based on SkyMapper DR1 \citep[][]{Wolf2018SkyMapper}. The excess of carbon, hence the exquisite strength of the CH $G$-band, is capable of depressing the continuum extending to the wavelength region of the \ion{Ca}{2} K/H lines, close to the center of SkyMapper's $v$ filter ($3825\angstrom$; see \citealt{DaCosta2019} and references therein), a phenomenon referred to as ``carbon veiling'' \citep[][]{yoon2020}. A scenario where, if confirmed, the surviving core of Sgr dSph has a lower CEMP fraction than its outskirts/stream at a given metallicity could be similar to what potentially happens to the Milky Way's bulge and halo \citep[][]{Arentsen2021pigs3}. Either way, the unbiased discovery of additional VMP stars in Sgr %\citep[e.g.,][]{Vitali2022SgrPristine} 
as well as other dSph satellites \citep[e.g.,][]{Skuladottir2021UMPsculptor} will be paramount for us to advance our understanding about the earliest stages of chemical enrichment in these systems.

Finally, we also calculate the fraction of CEMP stars in Sgr stream and compare it with the Milky Way. \citet[][]{Arentsen2022cemp} has recently demonstrated that various observational efforts focused on the discovery and analysis of metal-poor stars via low/medium-resolution (up to $\mathcal{R} \sim 3000$) spectroscopy %\footnote{The specific works analyzed were \citet[][]{lee2013}, \citet[][]{placco2018, placco2019}, \citet[][]{aguado2019}, \citet[][]{Arentsen2020pigs2}, \citet[][]{Yuan2020dtgs}, \citet[][]{Limberg2021_Gemini+SOAR}, and \citet[][]{Shank2022dtgs}.} 
report inconsistent CEMP fractions among them \citep{lee2013, placco2018, placco2019, aguado2019, Arentsen2020pigs2, Yuan2020dtgs, Limberg2021_Gemini+SOAR, Shank2022dtgs}. However, we reinforce that it is not our goal to provide absolute CEMP fractions \citep[e.g.,][]{rossi2005, lucatello2006, yoon2018}, but rather use the SEGUE/\texttt{StarHorse} low-metallicity sample to make a homogeneous comparison. For this reason, we do not perform any evolutionary corrections (as in \citealt{placco2014Carbon}) to the carbon abundances of \citet[][]{lee2013}. The overall fraction of CEMP stars in the whole sample at $\rm[Fe/H] < -2$, but excluding Sgr, is $19\pm1\%$\footnote{Uncertainties for fractions are given by Wilson score confidence
intervals \citep[][]{Wilson1927}. See \citet[][]{Limberg2021_Gemini+SOAR} for details.}  within the same $\log g$ range. For the whole Sgr stream, leading and trailing arms altogether, this number is $16\pm5\%$. Limiting our analysis to only giants ($T_{\rm eff} < 5800$\,K and $\log g < 4$), we find 12\% for both the stream and full sample. Therefore, we conclude the SEGUE carbon-abundance data does not provide evidence for variations in the CEMP frequency between Sgr (stream) and the Milky Way.

\section{Conclusions} \label{sec:conclusions}

In this work, we performed a chemodynamical study of Sgr stream, the tidal tails produced by the ongoing disruption of Sgr dSph galaxy. Because of recent literature results, we were particularly interested in exploring the VMP regime of this substructure. Our main goals were to quantify the kinematic properties of this population as well as search for CEMP stars. For the task, we leveraged low-resolution spectroscopic and astrometric data from SEGUE DR9 and \textit{Gaia} EDR3, respectively. Moreover, this catalog was combined with broad-band photometry from various sources in order to deliver %accurate and 
precise distances for ${\sim}175,000$ low-metallicity ($\rm[Fe/H] \leq -0.5$) stars via Bayesian isochrone-fitting in a new \texttt{StarHorse} run (Figure \ref{fig:xyz}), an effort that is fully described in an accompanying paper (Queiroz et al., submitted%, see also \citealt{Perottoni2022gse}
). Our main conclusions can be summarized as follows.

\raggedbottom

\begin{itemize}
    \item We delineated a new set of selection criteria for the Sgr stream based on %a combination between 
    angular momenta and actions (Figure \ref{fig:selection}). Despite being more conservative than previous works (e.g., \citetalias[][]{Johnson2020sgr} and \citealt[][]{naidu2020}), we identify ${\sim}1600$ members of Sgr stream, which is twice as many as these authors. Out of these, there are ${>}200$ VMP stars as well as 7 GCs. 
% (1) We delineated a new set of selection criteria for the Sgr stream based on a combination between angular momenta and actions. Despite being more conservative than previously considered by, e.g., J20 and Naidu+2020, we identify ~1600 members of Sgr stream, which twice as many as these authors' works. Out of these, there are >200 VMP stars as well as 7 GCs.

\item Reassuringly, although the SEGUE target selection inflates the number of metal-poor stars ($\rm[Fe/H] < -1$; Figure \ref{fig:sgr_sample}), we found the leading arm to be more metal-poor, by ${\sim}0.2$\,dex, than the trailing one. This is in agreement with many previous works \citep[notably][]{Hayes2020}.
% (2) Reassuringly, although the SEGUE target selection inflates the number of metal-poor stars ([Fe/h] < -1), we found the leading arm to be more metal-poor (by ~0.2 dex) than the trailing one. This is in agreement with many previous works (notably Hayes+2020).

\item We provided the first age estimates for individual stars in Sgr stream. For the task, we constructed a subsample of 56 turnoff/subgiant stars in this substructure, for which \texttt{StarHorse} ages are most reliable. We found an overall median age of $11.7^{+0.3}_{-0.2}\,{\rm Gyr}$, which is ${\sim}1$\,Gyr older than the bulk of thick-disk stars according both to our own SEGUE/\texttt{StarHorse} data as well as asteroseismic estimates \citep[][]{Miglio2021age_disks}.
% (3) We attempted to provide the first estimate of ages for individual stars in Sgr stream. For the task, we constructed a subsample of 56 turnoff/subgiant stars, for which StarHorse (i.e., isochrone-fitting) ages are most reliable. We found an overall median age of 11.7^{+0.3}_{-0.2}\,{\rm Gyr} Gyr for the leading and trailing arms combined. This is ~1 Gyr older than the bulk of the thick-disk stars according both to our own SEGUE/StarHorse data as well as asteroseismic estimates (Miglio+2021 and others)

\item We found ($2\sigma$) evidence for increasing velocity dispersion in Sgr stream between its metal-rich and VMP populations (Figure \ref{fig:vel_disp}/Table \ref{tab:vel_metal}). Similar findings were presented by \citetalias[][]{Johnson2020sgr}, but were contested by \citetalias[][]{Penarrubia2021sgr}% who argued that these authors' sample was highly contaminated by Milky Way interlopers
. Now, we reassert the former's findings with a $2{\times}$ larger sample and more rigorously-selected Sgr stream members (Figure \ref{fig:vel_disp}/Table \ref{tab:vel_metal}).
% (4) We found (2sigma) evidence for increasing velocity dispersion in Sgr stream from its metal-rich to VMP population. Similar findings had been previously presented by J20, but were contested by PP21 who argued that J20's sample was highly contaminated by Milky Way interlopers. Now, we reassert such findings with a 2x larger and more rigorously-selected Sgr stream sample (Figure 3/Table 1).

\item With the $N$-body model of \citetalias[][]{Vasiliev2021tango}, we found that the new wrap (composed of stars recently stripped; $t_{\rm strip} < 2\,{\rm Gyr}$) of Sgr stream preferentially contains metal-rich ($\rm[Fe/H] > -1.0$) stars. Conversely, the old wrap ($t_{\rm strip} > 2\,{\rm Gyr}$) is exclusively associated with metal-poor stars ($\rm[Fe/H] < -1.0$) in phase space. Hence, the increasing velocity dispersion with decreasing [Fe/H] is driven by the mixture between these components, i.e., larger fractions of the old wrap are found at lower metallicities, while the metal-rich population is only representative of the new wrap (Figure \ref{fig:sgr_coords}).
% (5) In comparison to the N-body model of V21, we found that the new wrap (tstrip < 2 Gyr) of Sgr stream preferentially contains metal-rich ([Fe/H] > -1.0) stars. Conversely, the old wrap is exclusively associated with metal-poor stars ([Fe/H] < -1.0). Therefore, the increasing velocity dispersion with decreasing [Fe/H] is a byproduct of the mixture between components, i.e., larger fractions of the old wrap are can be found at lower metallicities, while the metal-rich population is only representative of the new wrap (Figure 4).

\item Looking at the initial snapshot of the \citetalias[][]{Vasiliev2021tango} simulation, we found that stars that end up forming the old wrap are dynamically hotter %(larger velocity dispersion) 
and less centrally concentrated than those that compose the new wrap% (stripped later)
. With the understanding that the old wrap contains stars of lower metallicities% than the new one
, this implies that the outskirts of Sgr dSph, prior to disruption, were more metal-poor than is core regions, i.e., internal [Fe/H] variations in the galaxy. %The self-consistent reconstruction of such metallicity gradient and comparisons with surviving dSph galaxies \citep[see][]{Kirby2011MetalGrads} will be the topic of a forthcoming contribution.
% (6) Looking at the initial snapshot of the V21 simulation, we found that stars that end up forming the old wrap (stripped earlier) are dynamically hotter and less centrally concentrated than those that compose the new wrap (stripped later). With the understanding that the old wrap contains stars of lower metallicities than the new one, this implies that the outskirts of Sgr dSph prior to disruption were more metal-poor than is core regions, i.e., a metallicity gradient.
 
\item On the chemical-abundance front, SEGUE data allowed us to verify that the [$\alpha$/Fe] of Sgr stream decreases with increasing [Fe/H]. Most important, at a given metallicity, we ascertained that the median [$\alpha$/Fe] of Sgr stream is lower than GSE's, in conformity with other recent efforts \citep[][]{Hasselquist2021dwarf_gals, Horta2022haloSubs, Limberg2022gse}.
% (7) SEGUE abundance data allowed us to verify that the [alpha/Fe] of Sgr stream decreases with increasing [Fe/H]. Most important, at a given metallicity, we found that the median [alpha/Fe] of Sgr stream is lower than GSE's, in greement with recent efforts. (hasselquist+2021, limberg+2022)

\item We confidently (${>}3\sigma$) identify CEMP stars in Sgr stream. Also, its CEMP fraction %($16\pm5\%$) 
is compatible ($1\sigma$) with the overall SEGUE catalog% ($20\pm1\%$)
. Hence, we argue that the apparent lack of CEMP stars in Sgr dSph \citep[][and references therein]{Chiti2020sgr} could be associated with target-selection effects and/or small sample sizes. %Nevertheless, carbon-abundance information for larger samples of VMP stars across the whole Sgr system will be necessary to investigate this discrepancy between the stream and the remaining core of this dSph galaxy.
% (8) We confidently (>3sigma) identify CEMP stars in Sgr stream. Also, its CEMP fraction is indistinguishable from the overall SEGUE catalog. Hence, we argue that the apparent lack CEMP in Sgr dSph could be associated with target-selection effects. Nevertheless, carbon-abundance information larger samples of VMP stars across the whole Sgr system will be necessary to investigate this discrepancy between the stream and the remaining core of the galaxy.
\end{itemize}

% final remarks
% say how important deep spectroscopic surveys can be and how powerful of tool it is to combine them with gaia things
% also, emphasize the starhorse bayesian strategy
% usage of tailored n-body models
% points us towards the direction of trying to find additional vmp stars in sgr (as well as other disrupting dwarf galaxies. maybe antlia 2???)
This paper emphasizes how powerful the synergy between deep spectroscopy and astrometric data can be in our quest to unravel the outer Galactic halo. It also shows how crucial the fully Bayesian approach of \texttt{StarHorse} is for the task of deriving precise parameters (mainly distances) even for faint stars. In fact, the SEGUE/\texttt{StarHorse} catalog provides a glimpse of the scientific potential that will be unlocked by the next generation of wide-field surveys such as 4MOST, SDSS-V%/Milky Way Mapper
, and WEAVE. Finally, we reinforce the importance of tailored $N$-body models as fundamental tools for interpreting of the complex debris left behind by disrupted dwarf galaxies in the Milky Way's halo.

%\texttt{Astropy} \citep{astropy, astropy2018}, 
\software{\texttt{corner} \citep{corner2016}, \texttt{gala} \citep{gala2017}, \texttt{jupyter} \citep{jupyter2016}, \texttt{matplotlib} \citep{matplotlib}, \texttt{NumPy} \citep{numpy}, \texttt{pandas} \citep{pandasSoftware}, \texttt{SciPy} \citep{scipy}, \texttt{scikit-learn} \citep{scikit-learn}, \texttt{TOPCAT} \citep{TOPCAT2005}.
}

\begin{acknowledgments}

The authors thank the referee for a timely and constructive review which has contributed to this work. G.L. is indebted to Alex Ji, Ani Chiti, Felipe Almeida-Fernandes, Ting Li, and Vini Placco for discussions and suggestions that contributed to the original manuscript as well as Anke Arentsen who provided feedback on a preprint version of it. G.L. also thanks several authors who provided observational or simulation data, namely Alice Minelli, Amina Helmi, Emma Dodd, Khyati Malhan, Sergey Koposov, and Zhen Yuan. G.L. is particularly grateful toward Eugene Vasiliev, who readily provided the initial snapshot of the \citetalias[][]{Vasiliev2021tango} simulation as well as assistance with the model. Finally, G.L. thanks all those authors who made their observational and/or simulation data publicly available and are referenced throughout this work. G.L., H.D.P., S.R., J.A., and R.M.S. extend heartfelt thanks to all involved with the ``Brazilian Milky Way group meeting", namely Eduardo Machado-Pereira, Fabrícia O. Barbosa, H\'elio J. Rocha-Pinto, Lais Borbolato, Leandro Beraldo e Silva, and Yuri Abuchaim. 

G.L. acknowledges FAPESP (procs. 2021/10429-0 and 2022/07301-5). H.D.P. also thanks FAPESP
(procs. 2018/21250-9 and 2022/04079-0). S.R. thanks support from FAPESP (procs. 2014/18100-4 and 2015/50374-0), CAPES, and CNPq. J.A. acknowledges funding from the European Research Council (ERC) under the European Union’s Horizon 2020 research and innovation program (grant agreement No. 852839). R.M.S. acknowledges CNPq (Proc. 306667/2020-7). A.P.-V. acknowledges the DGAPA-PAPIIT grant IA103122. Y.S.L. acknowledges support from the National Research Foundation (NRF) of Korea grant funded by the Ministry of Science and ICT (NRF-2021R1A2C1008679). Y.S.L. also gratefully acknowledges partial support for his visit to the University of Notre Dame from OISE-1927130: The International Research Network for Nuclear Astrophysics (IReNA), awarded by the US National Science Foundation.

This work has made use of data from the European Space Agency (ESA) mission {\it Gaia} (\url{https://www.cosmos.esa.int/gaia}), processed by the {\it Gaia} Data Processing and Analysis Consortium (DPAC, \url{https://www.cosmos.esa.int/web/gaia/dpac/consortium}). Funding for the DPAC has been provided by national institutions, in particular the institutions participating in the {\it Gaia} Multilateral Agreement.

Funding for the Sloan Digital Sky Survey IV has been provided by the Alfred P. Sloan Foundation, the U.S. Department of Energy Office of Science, and the Participating Institutions. SDSS-IV acknowledges support and resources from the Center for High Performance Computing  at the University of Utah. The SDSS website is \url{www.sdss.org}. SDSS-IV is managed by the Astrophysical Research Consortium for the Participating Institutions of the SDSS Collaboration. 

\end{acknowledgments}

\begin{acknowledgments}
    This research has made use of the VizieR catalogue access tool, CDS, Strasbourg, France (\url{https://cds.u-strasbg.fr}). The original description of the VizieR service was published in \citet{VizieR2000}.
\end{acknowledgments}

\appendix

\section{Other Dwarf-galaxy Polar Streams} \label{sec:polar_streams}

In order to test if our Sgr stream selection %(Section \ref{subsec:selection}) 
is robust against the presence of other known dwarf-galaxy polar stellar streams \citep[see][]{Malhan2021lms1}, we assembled literature data for Cetus \citep[][originally found by \citealt{Newberg2009cetus}]{Yuan2019cetus}, Orphan \citep[][first described by \citealt{Belokurov2006Streams, Belokurov2007Orphan}]{Koposov2019orphan}, LMS-1/Wukong \citep[][also \citealt{naidu2020}]{Yuan2020lms1}, and Helmi streams \citep[][discovered by \citealt{helmi1999}]{OHare2020}. In this process, we made an effort to compile only members of streams found with automatic algorithms \citep[][]{myeongStreamsAndClumps, myeongShards, Yuan2018starGO}. The only exception is Orphan, whose members were selected based on the stream's track on the sky as well as proper motions. Note that, originally, \citet[][]{Yuan2020lms1} dubbed the LMS-1/Wukong substructure as ``low-mass stellar-debris stream'', hence the acronym. Almost at the same time, \citet[][]{naidu2020} identified a very similar dynamical group of stars in $(L_z, E)$ with the H3 survey and referred to it as ``Wukong''. For now, we %decided to 
keep both nomenclatures, similar to what several authors adopt for GSE. Apart from the $J_z > J_R$ condition, we followed \citet[][]{Naidu2022mzr} and \citet[][which builds on \citealt{koppelmanHelmi}, \citealt{Aguado2021}, and \citealt{Limberg2021dtgs}]{Limberg2021hstr} to apply further constraints to both Orphan and Helmi streams, respectively, in order to guarantee better purity for these samples. 

\begin{figure*}[pt!]
\centering
\includegraphics[width=1.\columnwidth]{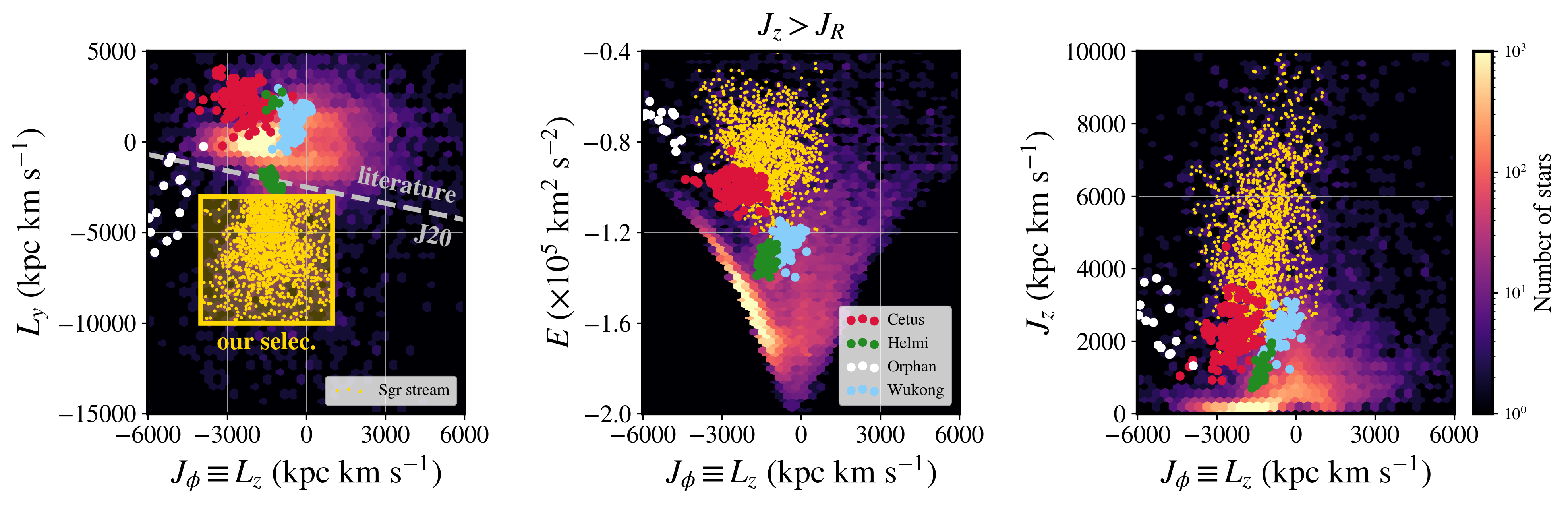}
\caption{Similar to the top row of Figure \ref{fig:selection} ($J_z > J_R$), but including literature data for dwarf-galaxy polar streams. Left: $(L_z, L_y)$. The dashed line shows the \citetalias[][]{Johnson2020sgr} criterion for selecting Sgr members. The yellow box exhibits our criteria. %The yellow cross marks the center of the Sgr stream locus in this parameter space according to \citetalias[][]{Penarrubia2021sgr}. Yellow contours illustrate the resulting kinematic/dynamical locus occupied by our Sgr stream stars. 
Colored dots represent stars from Sgr (yellow), Cetus (red), Orphan (white), LMS-1/Wukong (blue), and Helmi (green) streams (see text). Middle: $(L_z, E)$. Right: $(L_z, J_z)$ In all panels, density maps represent the full SEGUE/\texttt{StarHorse} low-metallicity sample. 
\label{fig:polar_streams}}
\end{figure*}

The kinematic/dynamical %\textit{volume} 
locus occupied by the polar streams is shown in Figure \ref{fig:polar_streams}. Crucially, they do not overlap the box in $(L_z, L_y)$ defined for Sgr (yellow region in the left panel). However, note how the \citetalias[][]{Johnson2020sgr} criteria actually encompasses the bulk of Orphan stream stars, which emphasizes the importance of our more rigorous selection. Helmi stream stars are the ones that reach the closest to our Sgr boundary in angular-momentum space. Although no stars from this substructure actually fulfill our entire set of selection criteria for Sgr, it is difficult to assert that Helmi stream interlopers are nonexistent in our Sgr sample. One way to quantify the cross-contamination between halo substructures would be to explore $N$-body models, similar to what has been recently done by \citet[][]{Sharpe2022halo}. For the time being, even without such a dedicated effort, we highlight that our criteria achieves several benchmarks, such as eliminating ${>}90\%$ of potential GSE stars, removing low-$J_z$ contaminants from the Galactic disk(s), and excluding most stars from other well-known dwarf-galaxy streams.

%Upper row: $J_z > J_R$ (predominantly polar orbits). Bottom: $J_z < J_R$ (radial/eccentric orbits). Background density maps are produced with the full SEGUE/\texttt{StarHorse} low-metallicity sample (Section \ref{subsec:starhorse}). White dots are Galactic GCs compiled by \citet{VasilievBaumgardt2021gcs}. Those GCs associated with Sgr dSph/stream (Section \ref{subsec:selection}) are displayed as yellow diamonds, with M54 as the star symbol. Left panels: $(L_z, L_y)$. Our Sgr stream selection is shown as the yellow box. The gray dashed line exhibits the \citetalias{Johnson2020sgr} criterion. The yellow cross is the central location of Sgr in this space according to \citetalias{Penarrubia2021sgr}. Yellow contours represent the dynamical locus occupied by our Sgr stream members. Middle: $(L_z, E)$. Right: $(L_z, J_z)$.

%\section{The presence of VMP stars on disk-like orbits} \label{sec:disk_vmp}

%\clearpage
\bibliography{bibliography.bib}{}
\bibliographystyle{aasjournal}

\end{document}